\documentclass{article}

\usepackage{arxiv}

\usepackage[utf8]{inputenc}
\usepackage[T1]{fontenc}
\usepackage{booktabs}
\usepackage{graphicx}
\usepackage[square,numbers]{natbib}
\usepackage{doi}
\usepackage{amsmath,amssymb,amsfonts}
\usepackage{physics}
\usepackage{bbm}
\usepackage{optidef}
\usepackage{algorithm}
\usepackage{algorithmicx}
\usepackage{algpseudocode}
\usepackage{multirow}
\usepackage{hyperref}

\hypersetup{colorlinks,allcolors=black}

\algrenewcommand{\Return}{\State\algorithmicreturn~}

\title{Bayesian Optimization of Sample Entropy Hyperparameters for Short Time Series}

\author{Zachary Blanks \\
	School of Data Science\\
	University of Virginia\\
	Charlottesville, VA 22903 \\
	\texttt{zdb6dz@virginia.edu} \\
	\And
    Donald E. Brown \\
	School of Data Science\\
	University of Virginia\\
	Charlottesville, VA 22903 \\
	\texttt{deb@virginia.edu} \\
}

\date{}

\renewcommand{\headeright}{}
\renewcommand{\undertitle}{}
\renewcommand{\shorttitle}{Bayesian Optimization of Sample Entropy Hyperparameters for Short Time Series}

\begin{document}
\maketitle

\begin{abstract}
Quantifying the complexity and irregularity of time series data is a primary pursuit across various data-scientific disciplines. Sample entropy (SampEn) is a widely adopted metric for this purpose, but its reliability is sensitive to the choice of its hyperparameters, the embedding dimension $(m)$ and the similarity radius $(r)$, especially for short-duration signals. This paper presents a novel methodology that addresses this challenge. We introduce a Bayesian optimization framework, integrated with a bootstrap-based variance estimator tailored for short signals, to simultaneously and optimally select the values of $m$ and $r$ for reliable SampEn estimation. Through validation on synthetic signal experiments, our approach outperformed existing benchmarks. It achieved a 60 to 90\% reduction in relative error for estimating SampEn variance and a 22 to 45\% decrease in relative mean squared error for SampEn estimation itself ($p \leq 0.043$). Applying our method to publicly available short-signal benchmarks yielded promising results. Unlike existing competitors, our approach was the only one to successfully identify known entropy differences across all signal sets ($p \leq 0.042$). Additionally, we introduce ``EristroPy,'' an open-source Python package that implements our proposed optimization framework for SampEn hyperparameter selection. This work holds potential for applications where accurate estimation of entropy from short-duration signals is paramount.
\end{abstract}


\section{Introduction}
\label{sec:introduction}
The ability to quantify the complexity or irregularity inherent in time series data is a primary pursuit across diverse data-scientific disciplines, including healthcare \citep{vikman1999altered, liang2021decreased, lake2002sample, lake2011accurate, wang2019sample, chan2021ultrasound}, finance \citep{yin2018complexity, olbrys2022regularity}, and anomaly detection systems \citep{zaman2018evaluation, virmani2016entropy, gilmary2021detection, huachun2021two, wang2020rolling}. Sample entropy (SampEn) has emerged as a widely adopted measure for this purpose \citep{richman2000physiological}.

SampEn estimates regularity through a process of ``template matching,'' partitioning the time series into overlapping segments or ``templates'' of length determined by the embedding dimension $(m)$. The similarity between templates is then evaluated based on their proximity within a defined radius $(r)$, typically measured in the $L_\infty$ space. A higher number of ``matches'' -- templates being within the radius $r$ of one another -- suggests greater regularity or predictability in the signal.

However, obtaining reliable SampEn estimates, particularly for short-duration signals, hinges on the careful selection of these two hyperparameters: $m$ and $r$ \citep{yentes2013appropriate}. 

Larger embedding dimensions $(m)$ are preferred to capture more complex time series patterns, but this tends to spread templates further apart in $L_\infty$ space, reducing the number of matches for a fixed radius $r$ \citep{kraemer2018recurrence}. Conversely, smaller values of $r$ allow more precise characterization of similarities between templates but also diminish the match count. This dynamic illustrates a fundamental trade-off: while larger $m$ and smaller $r$ are individually beneficial for characterizing signal complexity, their combined effect strains our ability to obtain a sufficient number of template matches required for stable SampEn estimation. Thus, the practical process of estimating SampEn for short-duration signals amounts to making deliberate trade-offs between picking larger values of $m$ and smaller values of $r$ to achieve stable SampEn estimates by ensuring more template matches.

\begin{figure*}
    \centering
    \includegraphics[width=\linewidth]{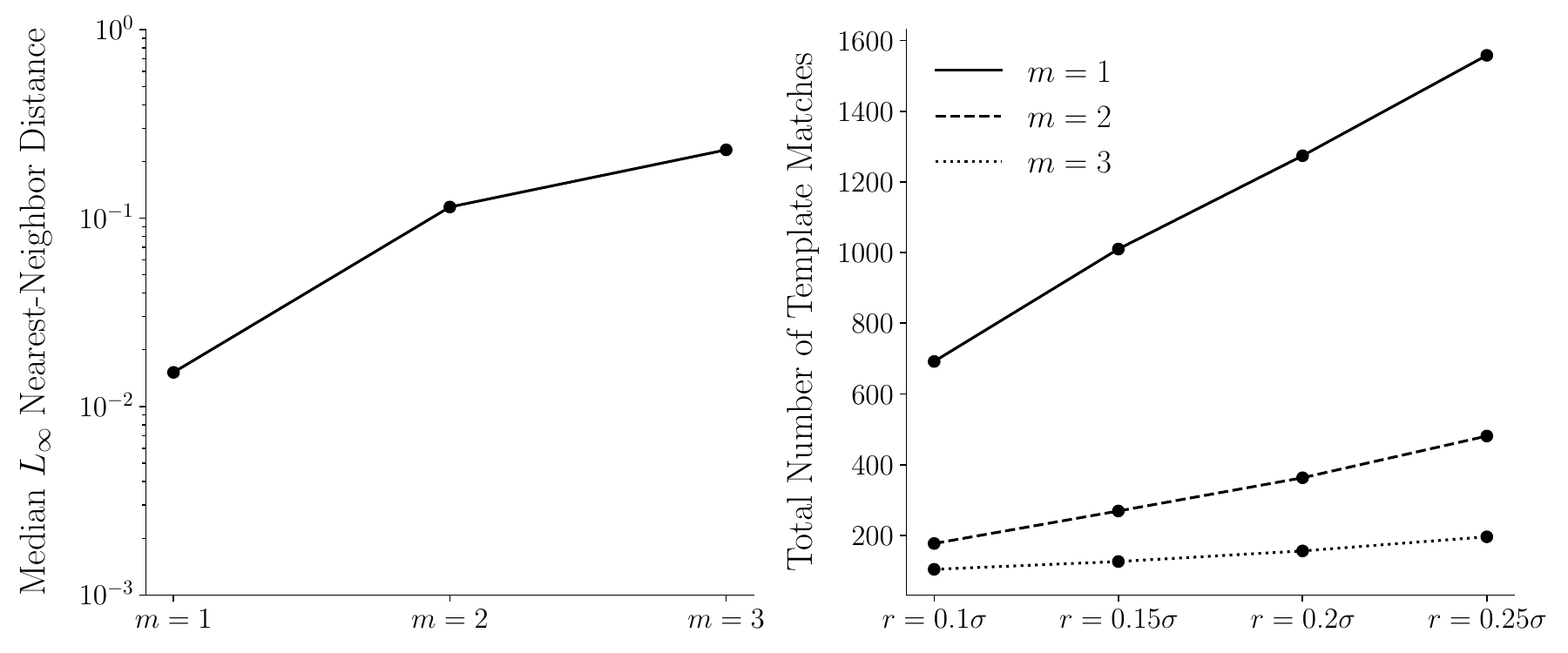}
    \caption{The interdependence of embedding dimension ($m$) and distance radius ($r$) illustrates a fundamental trade-off: as $m$ increases, templates diverge in $L_\infty$ space, reducing the number of matches for a fixed $r$, making it harder to obtain stable SampEn estimates. The multiplier $\sigma$ is set to one and denotes the signal standard deviation.}
    \label{fig:m-vs-r}
\end{figure*}

To illustrate this point, we depict a signal with $N = 100$ observations, computing the median $L_\infty$ distance between nearest neighbors among all $m$-dimensional templates and the total number of template matches for a fixed radius $r$. The results are shown in Fig. \ref{fig:m-vs-r} and empirically corroborate our previous claims. 

Existing approaches for selecting SampEn hyperparameters share a common limitation: the requirement to pre-determine the embedding dimension $m$, and then pick the similarity radius $r$ \citep{lake2002sample, ramdani2009use}. This strategy, however, neglects the interdependence between the hyperparameters. Furthermore, quantifying the variance of SampEn estimates obtained from short-duration signals, a necessary condition for appropriate hyperparameter selection, is particularly challenging due to the intra-correlated and non-linear construction of SampEn. Addressing these limitations is crucial for applications where accurate assessments of signal regularity are paramount, such as real-time clinical monitoring from short-duration physiological signals \citep{castaldo2019ultra, phinyomark2018feature, rahul2021short}.

The primary objective of this paper is to address the challenge of selecting $m$ and $r$ to achieve reliable SampEn estimates. We present two main contributions to advance the field of SampEn estimation for short-duration signals:

\begin{enumerate}
    \item A novel bootstrap-based, non-parametric estimator for SampEn variance, tailored for short signals, and a Bayesian optimization (BO) framework for the task of concurrently selecting optimal $(m, r)$ combinations.
    \item An open-source Python package, ``EristroPy,'' (\url{https://zblanks.github.io/eristropy/}) implementing our proposed hyperparameter selection algorithm, validated against existing benchmarks across diverse, publicly available signal datasets encompassing multiple signal types and measurement modalities.
\end{enumerate}

Through synthetic experiments and real-world case studies, we demonstrate the superiority of our approach in estimating SampEn variance and automating parameter selection over existing methods, consistently identifying known entropy differences between signal classes. Our work addresses a long-standing challenge in SampEn estimation, enabling more accurate and reliable assessments of signal complexity.

\section{Sample Entropy Overview}

\subsection{Sample Entropy Computation}
SampEn is a measure used to estimate the complexity or regularity of time series data. Developed by Richman and Moorman \citep{richman2000physiological}, it builds upon Pincus's ApEn \citep{pincus1991approximate}. Like related measures (e.g., Eckmann-Ruelle entropy \citep{eckmann1985ergodic}, ApEn, etc.), estimation of SampEn focuses on assessing regularity via ``template matching.''

Formally, let $\boldsymbol{x} \in \mathbb{R}^N$ be a time series signal of length $N$. A template of length $m$ (known as the ``embedding dimension'') is denoted by $\boldsymbol{x}_m^{(i)} = (x_i, \ldots, x_{i + m - 1})$. Implicitly, this definition of a template considers only consecutive data points from the signal. This does not have to be the case in general (formally captured through the time delay parameter, $\tau \in \mathbb{Z}_+$), but we fix $\tau = 1$ to ensure as large a sample size as possible, a common choice in the SampEn literature \citep{yentes2013appropriate, delgado2019approximate}.

We compare these templates and consider them a match if the distance between them is less than a defined radius, $r > 0$. Specifically, a match occurs when $\norm{\boldsymbol{x}_m^{(i)} - \boldsymbol{x}_m^{(j)}}_\infty \leq r$ ($i \neq j$), , denoting the maximum absolute difference between the elements of two templates. Unlike ApEn, SampEn does not consider self-matches -- i.e., $\norm{\boldsymbol{x}_m^{(i)} - \boldsymbol{x}_m^{(i)}}_\infty = 0$.

To calculate signal regularity, we compute two quantities, $B^m(r)$ and $A^m(r)$. $B^m(r)$, defined as:

\begin{equation}
    B^m(r) = \frac{1}{Z(N, m)} \sum_{i = 1}^{N-m} \sum_{\substack{j=1\\ j\neq i}}^{N-m} \mathbbm{1}\left[\norm{\boldsymbol{x}_m^{(i)} - \boldsymbol{x}_m^{(j)}}_\infty \leq r\right],
    \label{eq:sampen-bmr}  
\end{equation}

is the probability, in a frequentist sense, of the signal remaining within a radius, $r$, for $m$ points. Similarly, $A^m(r)$, given by:

\begin{equation}
    A^m(r) = \frac{1}{Z(N, m)}
    \sum_{i = 1}^{N-m} \sum_{\substack{j=1\\ j\neq i}}^{N-m} \mathbbm{1}\left[\norm{\boldsymbol{x}_{m+1}^{(i)} - \boldsymbol{x}_{m+1}^{(j)}}_\infty \leq r\right],
\label{eq:sampen-amr}
\end{equation}

is the probability that the signal stays within the same radius for an additional step ($m + 1$ points in total). In Eqns. (\ref{eq:sampen-bmr}) and (\ref{eq:sampen-amr}), $\mathbbm{1}\left[\cdot \right]$ denotes the indicator function and $Z(N, m)$ is a normalization constant to ensure valid probabilities and like-to-like comparisons between templates of size $m$ and $m+1$. For both $B^m(r)$ and $A^m(r)$ with fixed hyperparameters, $(m, r)$, their estimates become more accurate as the signal length, $N$, grows larger.

The ratio of these quantities: $\frac{A^m(r)}{B^m(r)}$ is the conditional probability (CP) that a sequence will remain within radius $r$ for $m + 1$ steps, given it has stayed within $r$ for $m$ steps. For fixed hyperparmeters $(m, r)$, the SampEn is defined as:

\begin{equation}
    \text{SampEn}(m, r) = \lim_{N \longrightarrow \infty} -\log\left( \frac{A^m(r)}{B^m(r)}\right).
\end{equation}

However, we of course do not have access to infinitely long signals and thus the best finite data estimation of SampEn for $\boldsymbol{x}$ is given by:

\begin{equation}
    \text{SampEn}(\boldsymbol{x}, m, r) = -\log\left( \frac{A^m(r)}{B^m(r)}\right).
    \label{eq:sampen-equation}
\end{equation}

If no matches are found at radius $r$ across the signal (i.e., $B^m(r) = 0$), $\text{SampEn}(\boldsymbol{x}, m, r)$ is undefined. Alternatively, if $B^m(r) > 0$ but $A^m(r) = 0$ then $\text{SampEn}(\boldsymbol{x}, m, r) = \infty$.

\subsection{Existing Approaches to Picking SampEn Hyperparameters}
Accurate and stable estimation of SampEn depends heavily on selecting appropriate values for the hyperparameters $(m, r)$. These choices significantly influence the computed SampEn value and, consequently, the interpretation of the data \citep{castiglioni2008threshold, liu2010comparison}. The prevalent guideline suggests setting $m = 2$ and $r \in [0.10, 0.25] \times \sigma$ -- the signal's standard deviation -- with $(m = 2, r = 0.20\sigma)$ -- often adopted as a standard \citep{delgado2019approximate}. This protocol, originally intended for ApEn in datasets with $N = 1000$ observations \citep{pincus1994physiological, pincus1991approximate}, has been empirically extended to smaller datasets, given SampEn's relative insensitivity to signal length \citep{delgado2019approximate, blanks2023signal}. Nonetheless, its applicability is questionable for very short time series (about $N \leq 200$ observations) \citep{yentes2013appropriate}. Recent studies in areas such as ultra-short heart rate variability \citep{castaldo2019ultra} and myoelectric sensor signals in prosthetics \citep{phinyomark2018feature} highlight these challenges.

One approach to circumvent the parameter selection challenge in SampEn estimation involves integrating it into a supervised machine learning classification framework, where SampEn serves as an input feature. Researchers have employed methods like grid-search over the parameters $m$ and $r$, in conjunction with various cross-validation techniques, to determine optimal parameter settings for the given data and classification models \citep{alcaraz2010optimal, cuesta2017noisy, cuesta2018model, padhye2022pressure}.

Nevertheless, this classification-focused methodology may not always be suitable. Often, the primary objective is to analyze signal complexity to gain insights into the dynamics of the physical process, rather than to differentiate between distinct classes. In such analyses, selecting SampEn parameters thoughtfully is crucial to accurately reflect the irregularity of the underlying system, rather than optimizing for classification accuracy. This paper will primarily focus on this aspect of SampEn application.

In contexts other than supervised learning, selecting appropriate parameters for SampEn remains an open challenge, particularly for the similarity radius parameter ($r$). Researchers have generally explored two main approaches outside the standard settings of $(m = 2, r = 0.20\sigma)$ for SampEn parameterization: optimization-based and convergence-based methods.

A key contribution in the optimization-based category, and a primary benchmark for this paper, is the method proposed by \citet{lake2002sample}. This approach involves selecting the radius that optimizes a SampEn efficiency criterion (SampEnEff) for a predetermined embedding dimension ($m$), usually fixed based on prior knowledge or an autoregressive analysis which finds the optimal time order lag determined via the Bayesian information criterion as a proxy for embedding dimension \citep{lake2002sample, schwarz1978estimating}. The objective function seeks to minimize the variability in SampEn estimates across different radius values:

\begin{equation}
     \min_{r \in \mathcal{R}} \quad \max\left(\frac{\sigma_{\text{CP}}}{\text{CP}}, \frac{\sigma_{\text{CP}}}{-\log(\text{CP})\text{CP}} \right).
    \label{eq:lake-obj}
\end{equation}

In Eq. (\ref{eq:lake-obj}), $\mathcal{R}$ denotes the set of feasible radius values, with $\text{CP}$ representing the conditional probability from the SampEn calculation (the ratio: $\frac{A^m(r)}{B^m(r)}$) and $\sigma_{\text{CP}}$ indicating the standard deviation of these estimates. The SampEnEff metric favors entropy estimates with lower variance. Their methodology was validated on neonatal heart rate data signals containing $N = 4096$ observations.

However, this approach may not be as effective with shorter time series. The assumptions underlying Eq. (\ref{eq:lake-obj}) that connects $\sigma_{\text{CP}}$ to the standard error (SE) of SampEn requires linearity in parameters and a Gaussian error distribution for CP estimation (predicated on the Central Limit Theorem). These assumptions might not be applicable to shorter signals due to the limited template match sample size. Consequently, the effectiveness of $\frac{\sigma_{\text{CP}}}{\text{CP}}$ as an approximation for the SE of SampEn in such cases is questionable.

Conversely, convergence-based strategies for SampEn parameter selection, originally introduced by \citet{ramdani2009use} and further developed or applied by \citet{kedadouche2015nonlinear} for SampEn and \citet{mengarelli2023multiscale} for fuzzy entropy \citep{chen2007characterization}, have gained traction recently. This rise in popularity may be attributed to their conceptual and computational simplicity. These methods, with the embedding dimension $m$ held constant, aim to identify a radius at which the SampEn estimates or their variance reach a stable point. Typically, the selection of $r$ is often conducted through visual analysis of a convergence plot, effectively seeking the ``elbow'' or ``knee'' point where increases in radius yield diminishing returns in reducing the variance of estimates. However, we formalize this ``knee''-finding approach using the algorithm detailed by \citet{satopaa2011finding}.

The efficacy of convergence-based approaches hinges on accurately determining the variance of SampEn estimates. In some scenarios, such as in the study by \citet{mengarelli2023multiscale}, simulating input signals simplifies the process of estimating variance. However, it is not universally feasible to model data as derivatives of simple synthetic signals \citep{liu2010comparison}. In these instances, either longer signals are needed to informally gauge variance, or one can employ the SampEn variance estimator proposed by \citet{lake2002sample} as a more formal approach.

\section{Methods}
\label{sec:methods}

\subsection{Mathematical Notations}
\label{subsection:notation}
We start by defining notations.

\begin{itemize}
    \item $\boldsymbol{x} \in \mathbb{R}^N$ is a time series signal containing $N$ observations. In general, we assume that $\boldsymbol{x}$ has zero mean and unit variance.
    \item $\boldsymbol{S} = \left\{\boldsymbol{x}_1, \ldots, \boldsymbol{x}_n \right\}$ is a collection of $n \in \mathbb{Z}_+$ signals where the $i$-th signal, $\boldsymbol{x}_i \in \mathbb{R}^{N_i}$, contains $N_i$ observations. We similarly assume that all the signals in $\boldsymbol{S}$ have zero mean and unit variance.
    \item $m \in \mathbb{Z}_+$ is the SampEn embedding dimension.
    \item $r \in \mathbb{R}_+$ is the SampEn similarity radius.
    \item $\hat{\theta}(m, r)$ denotes the SampEn estimate of $\boldsymbol{x}$ for a fixed ($m, r$) combination (see Eq. (\ref{eq:sampen-equation}))
    \item $B^m(r)$ and $A^m(r)$ are probabilities of the signal, $\boldsymbol{x}$, staying within $r$ for $m$ and $m+1$ points, respectively (see Eqns. (\ref{eq:sampen-bmr}) and (\ref{eq:sampen-amr})).
    \item $\text{CP} = \frac{A^m(r)}{B^m(r)}$ is the conditional probability that the signal, $\boldsymbol{x}$, stays within $r$ for $m+1$ points given that it has remained within $r$ for $m$ points.
    \item $q \in (0, 1)$ is the stationary bootstrap success probability characterizing the variable block sizes ($b \sim \text{Geom}(q))$ (see Section \ref{subsection:bootstrap-variance}).
    \item $\boldsymbol{x}_b \in \mathbb{R}^N$ is a bootstrapped signal of $\boldsymbol{x}$ constructed via Algorithm \ref{alg:stationary-bootstrap} given $q$.
    \item $\left\{\boldsymbol{x}_b \right\}_{b=1}^B$ is the collection of $B \in \mathbb{Z}_+$ bootstrapped signals given $q$.
    \item $\hat{\theta}_b(m, r)$ is the SampEn estimate obtained from the bootstrapped signal, $\boldsymbol{x}_b$, given $(m, r, q)$.
    \item $\left\{\hat{\theta}_b(m, r) \right\}_{b=1}^B$ is the collection of $B$ bootstrapped SampEn estimates obtained from the bootstrapped signal set, $\left\{\boldsymbol{x}_b \right\}_{b=1}^B$, given $(m, r, q)$.
    \item $\lambda \geq 0$ is a regularization control hyperparameter (see Eq. (\ref{eq:mse-optimization})).
    \item $\Psi_1 \subseteq \mathbb{Z}_+$, $\Psi_2, \Psi_3 \subseteq (0, 1)$ represent the domains of $m$, $r$, and $q$, respectively.
    \item $\Psi = \Psi_1 \times \Psi_2 \times \Psi_3$ is the domain of the decision variables, $\boldsymbol{\psi}$.
    \item $\boldsymbol{\psi} = (m, r, q) \in \Psi$ are the optimization decision variables.
    \item $f(\boldsymbol{\psi}) = \text{MSE}\left(\hat{\theta}(m, r)\right) + \lambda \sqrt{r}$ is the objective function for a signal signal, $\boldsymbol{x}$ (see Eqns. (\ref{eq:bootstrap-mse}) and (\ref{eq:mse-optimization})).
    \item $\tilde{f}(\boldsymbol{\psi})$ is the modified objective function for a signal set, $\boldsymbol{S}$ (see Eq. (\ref{eq:modified-obj})).
    \item $y_t = f(\boldsymbol{\psi}_t) + \epsilon_t$ represents the $t$-th observation of the objective function, $f: \Psi \rightarrow \mathbb{R}_+$, possibly observed with noise, $\epsilon_t$.
    \item $\mathcal{D} = \{(\boldsymbol{\psi}_t, y_t)\}_{t=1}^T$ is the set of observations.
    \item $\mathcal{D}^{(l)}$ and $\mathcal{D}^{(g)}$ are the sets of observations derived from $\mathcal{D}$ containing the better-performing and worse-performing objective function evaluations, respectively.
    \item $\gamma \in (0, 1]$ is the top quantile used to construct $\mathcal{D}^{(l)}$.
    \item $y^{\gamma} \in \mathbb{R}_+$ is the top-$\gamma$ objective value in $\mathcal{D}$.
    \item $p\left(\boldsymbol{\psi} \mid \mathcal{D}^{(l)}\right)$ and $p\left(\boldsymbol{\psi} \mid \mathcal{D}^{(g)}\right)$ are the probability density functions (PDFs) of the better-performing and worse-performing groups, respectively.
\end{itemize}

\subsection{Bootstrap-Based Variance Estimation for SampEn}
\label{subsection:bootstrap-variance}

Given a short-duration signal $\boldsymbol{x} \in \mathbb{R}^N$ (e.g., $N \leq 200$), our aim is to select SampEn hyperparameters $(m, r)$ to ensure we obtain a stable estimate of $\text{SampEn}(\boldsymbol{x}, m, r)$. For notational brevity, let $\hat{\theta}(m, r)$ denote the SampEn estimate of $\boldsymbol{x}$ for fixed $(m, r)$.

To ensure SampEn estimate stability, we require an accurate quantification of variance. We propose a bootstrap-based approach for this purpose. The bootstrap, introduced by Efron \citep{efron1992bootstrap}, generates empirical distributions of arbitrary statistics by resampling data. For SampEn, we resample $\boldsymbol{x}$, yielding a new signal, $\boldsymbol{x}_b \in \mathbb{R}^N$, which we use to calculate the bootstrapped SampEn estimate $\hat{\theta}_b(m, r)$. This process is repeated $B \in \mathbb{Z}_+$ times, giving the set: $\left\{\hat{\theta}_b(m, r) \right\}_{b=1}^B$.

Standard bootstrapping, however, assumes samples from the data are independent, an inappropriate assumption for time series data. We address this issue by using the stationary bootstrap \citep{politis1994stationary}. The main intuition behind this algorithm is that adjacent elements of the signal -- e.g., $(x_i, \ldots x_{i+b-1})$ -- should be treated as connected ``blocks.'' We then build the new bootstrapped signal, $\boldsymbol{x}_b$, by stitching randomly selected blocks together. Algorithm \ref{alg:stationary-bootstrap} details this method.

\begin{algorithm*}[htbp]
\caption{Stationary Bootstrap Signal Generation}
\begin{algorithmic}[1]
    \Require $\boldsymbol{x} \in \mathbb{R}^N$, $N \in \mathbb{Z}_+$, $q \in (0, 1)$
    \State $B \gets []$ \Comment{Placeholder for data blocks from $\boldsymbol{x}$}
    \State $l \gets 1$
    \While{$l \leq N$}
        \State $t_0 \gets \mathcal{U}\left(1, \ldots, N \right)$
        \State $b \gets \text{Geom}(q)$
        \State $t_1 \gets t_0 + b - 1$ \Comment{Handle $t_1 > N$ by ``wrapping'' the index around}
        \If{$l + b > N$} \Comment{We cannot construct bootstrapped signals greater than $N$ observations}
            \State $\tilde{b} \gets b - (l + b - N)$
            \State $t_1 \gets t_0 + \tilde{b} - 1$
        \EndIf
        \State $B \gets \text{append}\left(\left[x_{t_0}, \ldots, x_{t_1} \right] \right)$
        \State $l \gets l + b$
    \EndWhile
    \State $\boldsymbol{x}_b \gets$ concatenate the blocks in $B$
    \Return $\boldsymbol{x}_b$
\end{algorithmic}
\label{alg:stationary-bootstrap}
\end{algorithm*}

There are two noteworthy points. First, the stationary bootstrap uses variable block sizes ($b \sim \text{Geom}(q)$ for $q \in (0, 1)$) to mitigate overall sensitivity to block size \citep{politis2004automatic}. Second, to handle the stop index, $t_1$, exceeding signal length, we wrap indices around. That is, if $t_1 = N + a$ where $a > 0$ then the elements we use to construct the data block are: $\left[x_{t_0}, \ldots, x_N, x_1, \ldots, x_{1 + a - 1}  \right]$. 

With a fixed $q \in (0, 1)$, and bootstrapped SampEn estimates $\left\{\hat{\theta}_b(m, r) \right\}_{b=1}^B$, we compute the variance of the SampEn estimate as:

\begin{equation}
    \mathbb{V}\left(\hat{\theta}(m, r) \right) \approx \frac{1}{B} \sum_{b=1}^B \left(\hat{\theta}_b(m, r) - \bar{\theta}(m, r) \right)^2,
    \label{eq:bootstrap-variance}
\end{equation}

where $\bar{\theta}(m, r)$ is the mean of bootstrapped SampEn estimates.

\subsection{Minimizing the SampEn Estimate Mean Squared Error}
\label{section:sampen-param-selection}

Recall that $\left\{\hat{\theta}_b(m, r) \right\}_{b=1}^B$ is the set of $B$ SampEn estimates obtained from the stationary bootstrap replicates of $\boldsymbol{x}$ for a fixed $q \in (0, 1)$ and SampEn hyperparameters $(m, r)$.

We seek a stable SampEn estimate of $\boldsymbol{x}$ by manipulating the hyperparameters $(m, r, q)$, aiming for low bias -- the systematic deviation of the bootstrap estimates from the original SampEn estimate, $\hat{\theta}(m, r)$ -- and low variance -- sensitivity of the bootstrap estimates to minor signal variations.

The bias of $\hat{\theta}(m, r)$ is approximated by the expression:

\begin{equation}
    \text{Bias}\left(\hat{\theta}(m, r) \right) \approx \left(\frac{1}{B}\sum_{b=1}^B \hat{\theta}_b(m, r)\right) - \hat{\theta}(m, r).
\end{equation}

Conveniently, the mean squared error (MSE) of a statistic -- a common loss function -- corresponds to the squared bias plus the variance of that statistic. Thus, the approximate MSE of $\hat{\theta}(m, r)$ is given by:
    
\begin{align}
    \begin{aligned}
        \text{MSE}\left(\hat{\theta}(m, r) \right) &= \left(\text{Bias}\left(\hat{\theta}(m, r)\right) \right)^2 + \mathbb{V}\left(\hat{\theta}(m, r) \right) \\
        &\approx \frac{1}{B}\sum_{b=1}^B \left(\hat{\theta}(m, r) - \hat{\theta}_b(m, r) \right)^2.
    \end{aligned}
    \label{eq:bootstrap-mse}
\end{align}

To then select the optimal values of $(m, r, q)$ which minimizes the approximate MSE of $\hat{\theta}(m, r)$, we compute:

\begin{mini}
    {m, r, q}{\text{MSE}\left(\hat{\theta}(m, r) \right) + \lambda \sqrt{r}}{\label{eq:mse-optimization}}{}
    \addConstraint{m}{\in \mathbb{Z}_+}{}
    \addConstraint{r, q}{\in (0, 1)}{}.
\end{mini}

We avoid trivial solutions from Eq. (\ref{eq:mse-optimization}) where $r \rightarrow \infty$ by introducing the regularization function, $\Omega(r) = \lambda \sqrt{r}$ and constrain $r \in (0, 1)$ where $\lambda \geq 0$ acts as the regularization parameter and assume a normalized signal with zero mean and unit variance. Values for $r > \sigma$, where $\sigma$ is the empirical standard deviation of $\boldsymbol{x}$, are likely not a meaningful characterization of similarity for short-duration signals and would notably diverge from standard recommendations \citep{delgado2019approximate}. Furthermore, over $(0, 1)$, $\sqrt{r}$ provides a greater degree of regularization than the more canonical $L_1$ regularization: $\abs{r}$, and the $L_2^2$ regularization: $r^2$.

Problem (\ref{eq:mse-optimization}) is difficult to solve because the objective function is non-convex and we do not have access to its gradients due to our proposed bootstrapping procedure.

\subsection{BO-driven SampEn Hyperparameter Search}
\label{section:bayes-opt}
Consider the problem $\min_{\boldsymbol{\psi} \in \Psi} \ f(\boldsymbol{\psi})$, where $\boldsymbol{\psi}$ represents decision variables, $\Psi$ is their constraint set or domain, and $f$ is a continuous, non-convex, and ``black-box’’ objective function. In our case, this problem corresponds to selecting optimal SampEn hyperparameters to achieve stable estimates

Bayesian optimization (BO) is well-suited for such problems, finding locally optimal solutions efficiently \citep{frazier2018bayesian}. We employ the Tree-structured Parzen Estimator (TPE) method \citep{bergstra2011algorithms}, known for its effectiveness in similar hyperparameter optimization tasks \citep{watanabe2023tree}.

BO iteratively explores the decision space, $\Psi$, intelligently selecting new points for evaluation based on previous observations and uncertainties about unexplored regions.

\begin{figure*}
    \centering
    \includegraphics[width=0.9\linewidth]{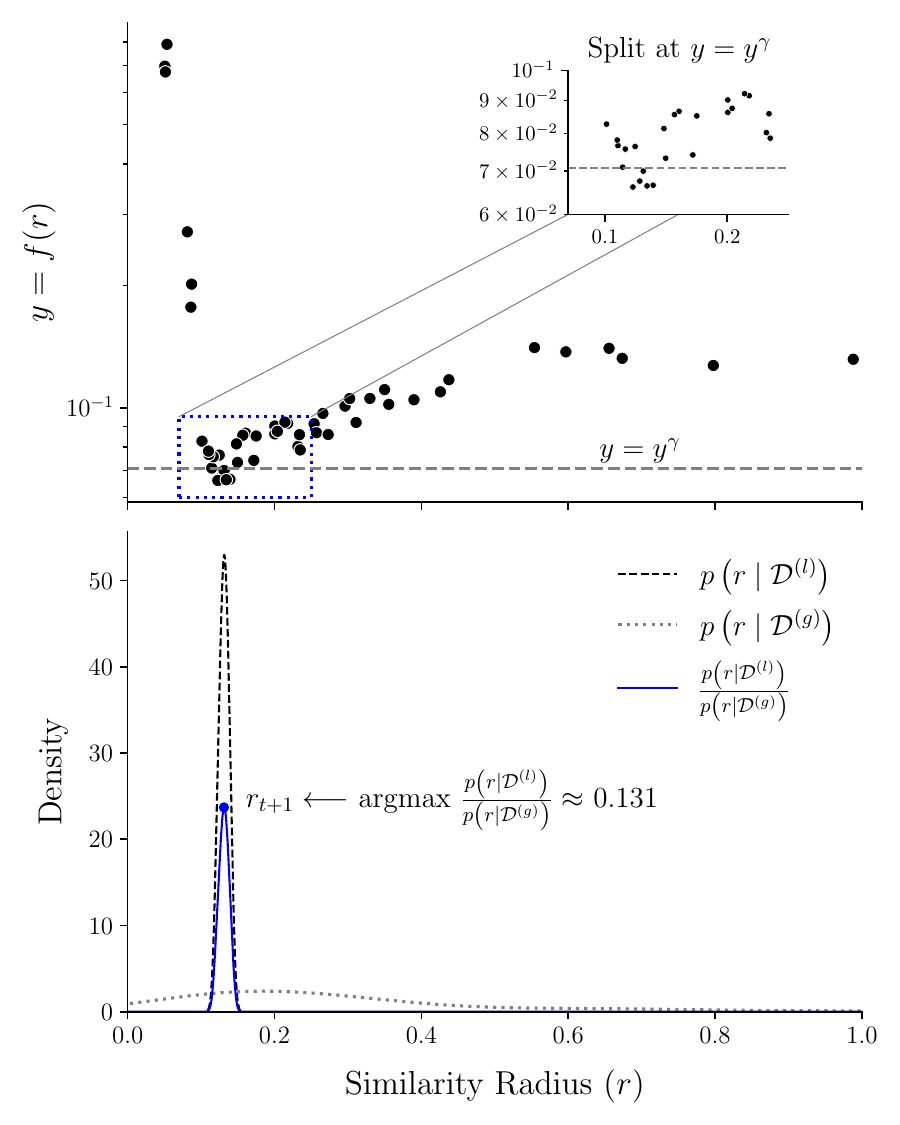}
    \caption{Simplified depiction of the Tree-structured Parzen Estimator (TPE) process for choosing a new evaluation point during the optimization procedure. (Top): We partition the objective function evaluations into better and worse-performing sets using a threshold cutoff, $y \leq y^\gamma$. (Bottom): We construct the probability densities for the better and worse-performing sets, then select the new evaluation point as the argmax of the ratio of these densities.}
    \label{fig:tpe-process}
\end{figure*}

In the TPE method, shown schematically for a simplified example focusing on the similarity radius, $r$, in Fig. \ref{fig:tpe-process}, we partition past objective function evaluations into better and worse-performing sets using a threshold cutoff, $y^{\gamma} \in \mathbb{R}_+$. Then, we construct probability densities for these sets and select the next evaluation point as the argmax of the ratio of these densities.

To apply the TPE method to the SampEn hyperparameter selection problem, we start by considering how to select the $(T+1)$-th evaluation point given $T$ previous objective function evaluations. We summarize this process in Algorithm \ref{alg:tpe-new-point}.

\begin{algorithm*}[htbp]
\caption{TPE New Point Selection Process}
\begin{algorithmic}[1]
    \Require $T$ previous objective function evaluations, $\mathcal{D} = \left\{\left(\boldsymbol{\psi}_t, y_t \right) \right\}_{t=1}^T$, $S \in \mathbb{Z}_+$
    \State Compute top-quantile cut-off, $\gamma = \Gamma(T)$ \Comment{See Appendix \ref{subsection:top-quantile-selection}}
    \State Sort $\mathcal{D}$ by $y_t$ such that $y_1 \leq y_2 \leq \cdots \leq y_T$
    \State Compute $T^{(l)} = \min\left(\lceil \gamma T \rceil, 25\right)$
    \State Construct $\mathcal{D}^{(l)} = \left\{(\boldsymbol{\psi}_t, y_t) \right\}_{t=1}^{T^{(l)}}$ and $\mathcal{D}^{(g)} = \left\{(\boldsymbol{\psi}_t, y_t) \right\}_{t=T^{(l)} + 1}^T$
    \State Compute $\left\{w_t \right\}_{t=0}^{T+1}$ \Comment{See Appendix \ref{subsection:gaussian-weights}}
    \State Build $p\left(\boldsymbol{\psi} \mid \mathcal{D}^{(l)} \right)$ and $p\left(\boldsymbol{\psi} \mid \mathcal{D}^{(g)} \right)$ \Comment{See Appendix \ref{subsection:kernel-and-bandwidth} and \ref{subsection:surrogate-model}}
    \State Sample $\mathcal{S} \equiv \left\{\boldsymbol{\psi}_s \right\}_{s=1}^S \sim p\left(\boldsymbol{\psi} \mid \mathcal{D}^{(l)} \right)$
    \State $\boldsymbol{\psi}_{T+1} \gets \text{argmax}_{\boldsymbol{\psi} \in \mathcal{S}} \ \frac{p\left(\boldsymbol{\psi} \mid \mathcal{D}^{(l)} \right)}{p\left(\boldsymbol{\psi} \mid \mathcal{D}^{(g)} \right)}$
    \Return $\boldsymbol{\psi}_{T+1}$
\end{algorithmic}
\label{alg:tpe-new-point}
\end{algorithm*}

Suppose we have $T$ observations obtained by evaluating $f$ as $\mathcal{D} = \{(\boldsymbol{\psi}_t, y_t)\}_{t=1}^T$ and assume these observations are sorted by $y_t$ such that $y_1 \leq y_2 \leq \cdots \leq y_T$. Let $\gamma \in (0, 1]$ be the top quantile of the observations, and $y^{\gamma}$ be the top-$\gamma$ objective value in $\mathcal{D}$. We partition $\mathcal{D}$ into a better-performing set, $\mathcal{D}^{(l)}$, and worse-performing set, $\mathcal{D}^{(g)}$, where $\mathcal{D}^{(l)} = \{(\boldsymbol{\psi}_t, y_t)\}_{t=1}^{T^{(l)}}$ and $T^{(l)} = \min\left(\lceil \gamma T \rceil, 25 \right)$, and $\mathcal{D}^{(g)}$ contains the remaining observations.

Under the TPE method, the surrogate model is the conditional probability density function:

\begin{equation}
    p\left(\boldsymbol{\psi} \mid y, \mathcal{D} \right) = \begin{cases}
        p\left(\boldsymbol{\psi} \mid \mathcal{D}^{(l)} \right), &\text{if, } y \leq y^{\gamma} \\
        p\left(\boldsymbol{\psi} \mid \mathcal{D}^{(g)} \right), &\text{if, } y > y^\gamma \\
    \end{cases}
\end{equation}

and the acquisition function to select new evaluation points is defined as:

\begin{equation}
    \boldsymbol{\psi}_{T+1} \coloneqq \text{argmax}_{\boldsymbol{\psi} \in \Psi} \ \frac{p\left(\boldsymbol{\psi} \mid \mathcal{D}^{(l)} \right)}{p\left(\boldsymbol{\psi} \mid \mathcal{D}^{(g)} \right)}.
\end{equation}

The TPE method diverges from standard BO by modeling how decision variables, $\boldsymbol{\psi}$, are affected by function evaluations, rather than modeling the objective function itself \citep{bergstra2011algorithms}. This strategy uses the intuition that promising evaluation points are likely found within regions containing higher-performing observations. We construct the PDFs, $p(\boldsymbol{\psi} \mid \mathcal{D}^{(l)})$ and $p(\boldsymbol{\psi} \mid \mathcal{D}^{(g)})$ using a mixture of truncated Gaussian kernel density estimators (KDEs).

The surrogate model involves several hyperparameter choices, including $\gamma$, the kernel estimator, $k$, the bandwidth parameter, $b$, and mixture weights, $\left\{w_t \in [0, 1] \right\}_{t=0}^{T+1}$. We adopt default values from the ``Optuna" optimization framework \citep{optuna_2019}, based on prior research (e.g., \citep{bergstra2011algorithms, bergstra2013making, watanabe2023tree, song2022general}). Further details about these hyperparameters are provided in Appendix \ref{appendix:tpe-hyperparams}.

\subsection{SampEn Hyperparameter Optimization Algorithm}
\label{section:solving-the-problem}

We present Algorithm \ref{alg:sampen-bayes-opt} to outline our approach for finding a locally optimal solution to the SampEn hyperparameter selection problem for the signal, $\boldsymbol{x} \in \mathbb{R}^N$. The decision variables $\boldsymbol{\psi} = (m, r, q) \in \Psi$, representing the SampEn hyperparameters $(m, r)$ and the stationary bootstrap parameter, $q$, are optimized with respect to the objective function, $f(\boldsymbol{\psi})$ (defined in Section \ref{subsection:notation}).

\begin{algorithm*}[htbp]
\caption{SampEn Hyperparameter Selection Optimization}
\begin{algorithmic}[1]
\Require $\boldsymbol{x} \in \mathbb{R}^N$, $B \in \mathbb{Z}_+$, $\widetilde{T} =$ Number of BO trials, $\lambda \geq 0$, $T_{\text{init}} \in \mathbb{Z}_+$
\State $\mathcal{D} \gets \emptyset$ \Comment{Empty observation set}
\For{$t = 1, \ldots, T_{\text{init}}$}
    \State Select $\boldsymbol{\psi}_t$ randomly
    \State Compute SampEn estimate of $\boldsymbol{x}$, $\hat{\theta}(m_t, r_t)$
    \State Generate bootstrap replicates of $\boldsymbol{x}$, $\left\{\boldsymbol{x}_b \right\}_{b=1}^B$, given $q_t$ \Comment{See Algorithm \ref{alg:stationary-bootstrap}}
    \State Compute bootstrap SampEn estimates, $\left\{\hat{\theta}_b(m_t, r_t) \right\}_{b=1}^B$
    \State $y_t = f\left(\boldsymbol{\psi}_t \right) = \text{MSE}\left( \hat{\theta}(m_t, r_t)\right) + \lambda \sqrt{r_t}$
    \State $\mathcal{D} \gets \mathcal{D} \cup \left(\boldsymbol{\psi}_t, y_t \right)$
\EndFor
\While{$t \leq \widetilde{T}$}
    \State Select $\boldsymbol{\psi}_{t+1}$ using the TPE acquisition process \Comment{See Algorithm \ref{alg:tpe-new-point}}
    \State Compute SampEn estimate, $\hat{\theta}(m_{t+1}, r_{t+1})$
    \State Generate bootstrap replicates of $\boldsymbol{x}$, $\left\{\boldsymbol{x}_b \right\}_{b=1}^B$, given $q_{t+1}$ \Comment{See Algorithm \ref{alg:stationary-bootstrap}}
    \State Compute bootstrap SampEn estimates, $\left\{\hat{\theta}_b(m_{t+1}, r_{t+1}) \right\}_{b=1}^B$
    \State $y_{t+1} = f\left(\boldsymbol{\psi}_{t+1} \right) = \text{MSE}\left( \hat{\theta}(m_{t+1}, r_{t+1})\right) + \lambda \sqrt{r_{t+1}}$
    \State $\mathcal{D} \gets \mathcal{D} \cup \left(\boldsymbol{\psi}_{t+1}, y_{t+1} \right)$
\EndWhile
\State $y^* \gets \min \ \left\{y_1, \ldots, y_{\widetilde{T}} \right\}$
\State $\boldsymbol{\psi}^* \gets \text{argmin} \ \left\{y_1, \ldots, y_{\widetilde{T}} \right\}$
\Return $\boldsymbol{\psi}^*, y^*$
\end{algorithmic}
\label{alg:sampen-bayes-opt}
\end{algorithm*}

One key aspect of Algorithm \ref{alg:sampen-bayes-opt} is the random initialization phase, where we start the search process using $T_{\text{init}} \in \mathbb{Z}_+$ replicates. Throughout this manuscript, we adopt the default value provided by Optuna \citep{optuna_2019}, $T_{\text{init}} = 10$. Following the initial $T_{\text{init}}$ evaluations, we transition to employing the TPE method, as explained in Algorithm \ref{alg:tpe-new-point}, to iteratively explore the decision space, $\Psi$, and selecting new evaluation points.

\subsection{Optimizing SampEn Hyperparameters for Multi-signal Analysis}
In SampEn-based analyses involving multiple signals, it is customary to select a fixed $(m, r)$ combination for all signals to ensure consistent comparisons. While our algorithm thus far has focused on optimizing $(m, r)$ for a single signal $\boldsymbol{x} \in \mathbb{R}^N$, extending it to accommodate a signal set $\boldsymbol{S} = \left\{\boldsymbol{x}_1, \ldots, \boldsymbol{x}_n \right\}$ with $n \in \mathbb{Z}_+$ signals is straightforward.

Algorithm \ref{alg:sampen-bayes-opt-multiple-signals} in Appendix \ref{appendix:signal-set-opt-algo} outlines this extension. For each signal, $\boldsymbol{x}_i$, we compute the SampEn estimate, $\hat{\theta}_i(m, r)$, generate $B$ bootstrap replicates, $\left\{\boldsymbol{x}_{i,b} \right\}_{b=1}^B$, and calculate the bootstrapped SampEn estimates, $\left\{\hat{\theta}_{i, b}(m, r) \right\}_{b=1}^B$. Using these, we estimate the SampEn MSE of $\boldsymbol{x}_i$ given $(m, r, q)$. The modified objective function, $\tilde{f}(\boldsymbol{\psi})$,  accounts for $n$ signals and is defined as:

\begin{equation}
    \tilde{f}(\boldsymbol{\psi}) =\left(\frac{1}{n} \sum_{i=1}^n \text{MSE}\left(\hat{\theta}_i(m, r) \right) \right) + \lambda\sqrt{r}.
    \label{eq:modified-obj}
\end{equation}

Under this scheme, the process to select new evaluation points using the TPE method remains unchanged. We simply use the mean of the regularized MSE estimates across the signal set, $\boldsymbol{S}$, to select a unified value of $(m, r)$.

\section{Results}
All computational experiments were performed with a MacBook Air equipped with an Apple M1 chip with 16GB of RAM running macOS Sonoma 14.4.1, using Python 3.11. We developed a Python package,  ``EristroPy,'' which provides a complete implementation of our proposed SampEn hyperparmaeter selection strategy. Detailed documentation for ``EristroPy" can be found at \url{https://zblanks.github.io/eristropy}.

\subsection{Synthetic Signal Experiments}
We assessed the performance of our proposed SampEn hyperparameter optimization algorithm using synthetic white noise and autoregressive order one (AR(1)) signals. Our evaluation focused on the following aspects:

\begin{itemize}
    \item The accuracy of our proposed SampEn variance estimator (Eq. \ref{eq:bootstrap-variance}) compared to the estimator by \citet{lake2002sample}.
    \item The impact of the regularization parameter $\lambda$ on the optimization procedure.
    \item The search behavior of the BO algorithm within the $(m, r, q)$ decision space.
    \item The influence of the stationary bootstrap success probability $q$.
    \item The overall SampEn hyperparameter selection performance relative to existing benchmarks.
    \item The computational efficiency of the proposed method.
\end{itemize}

White noise is a series of independent, identically distributed random variables: $x_t \sim \mathcal{N}(0, \sigma^2)$ for $\sigma > 0$. AR(1) processes are stochastic models where each value in the series is linearly dependent on the previous value plus a random noise term: $x_{t} = \phi x_{t-1} + \epsilon_{t}$ where $\epsilon_t \sim \mathcal{N}(0, \sigma^2)$ for $\sigma > 0$ and $\phi \in \mathbb{R}$. The AR(1) signal class, thus, represents a signal containing both stochastic and deterministic components.

For our experiments, we set $\sigma = 1$ for white noise and $\phi = 0.9$ with $\sigma = 0.1$ for AR(1) processes. We ensured statistical signal stationarity according to the Augmented Dickey-Fuller (ADF) test \citep{dickey1979distribution} by applying a burn-in period of $500$ samples for AR(1) processes.

\subsubsection{Evaluating SampEn Variance Estimators for Short Signals}
\label{section:variance-estimation}
We hypothesized that the SampEn variance estimator proposed by \citet{lake2002sample}, relying on template match counting and uncertainty propagation, might not be optimal for short signal settings. To test this, we compared our bootstrap-based estimator with Lake et al.'s estimator across white noise and AR(1) signal classes, varying the signal length $N$ and similarity radius $r$, with fixed $m = 1$ and different $q$ values for each signal type. We chose $q = 0.5$ for the AR(1) signal class to yield an expected block size of $b = 2$ aligning with its time-order dependence, and $q = 0.9$ for Gaussian white noise, accounting for its lack of temporal dependence.

The mean percent difference in SampEn variance estimation error, $\Delta \epsilon(r, N)$, was assessed for $N \in \{50, 100, 200\}$ and $r \in \{0.20\sigma, 0.25\sigma\}$. Positive values of $\Delta \epsilon(r, N)$ indicate our bootstrap-based estimator outperforms Lake et al.'s approach in terms of SampEn variance estimation error. Summary statistics of $\Delta \epsilon(r, N)$ are presented in Table \ref{tab:var-comparison}, showing consistent superiority of our proposed method across all signal types, similarity radii, and signal lengths. Refer to Appendix \ref{appendix:sampen-variance-experiment-details} for experimental and computational specifics.

\begin{table*}[htbp]
\centering
\caption{Our proposed bootstrap-based SampEn variance estimator demonstrates statistically lower error rates compared to the ``counting''-based approach by \protect\citet{lake2002sample}. Across signal type, $N$, and $r$ combinations, we achieve a reduction in SampEn variance estimation mean squared error typically ranging from 60\% to 90\% relative to the counting approach. The $\overline{\Delta \epsilon}(r, N)$ column is the mean posterior estimate for the relative percentage difference in SampEn variance estimation error (see Eq. (\ref{eq:variance-comparison-model}) in Appendix \ref{appendix:sampen-variance-experiment-details}) at similarity radius $r$ and signal length $N$.}
\label{tab:var-comparison}
\begin{tabular}{ccccc}
\toprule
Signal Type & $N$ & $r$          & $\overline{\Delta \epsilon}(r, N)$ & 94\% Credible Interval         \\
\midrule
\multirow{6}{*}{White Noise} & \multirow{2}{*}{50}  & $0.20\sigma$ & 69.16                      & [68.36,70.08] \\
                             &     & $0.25\sigma$ & 79.44                      & [78.99,79.88] \\ 
                             & \multirow{2}{*}{100} & $0.20\sigma$ & 90.15                      & [89.95,90.36] \\ 
                             &     & $0.25\sigma$ & 91.46                      & [91.29,91.64] \\ 
                             & \multirow{2}{*}{200} & $0.20\sigma$ & 94.42                      & [94.32,94.52] \\ 
                             &     & $0.25\sigma$ & 94.93                      & [94.84,95.01] \\
\\
\multirow{6}{*}{AR(1)}       & \multirow{2}{*}{50}  & $0.20\sigma$ & 64.04                      & [63.71,64.37] \\ 
                             &     & $0.25\sigma$ & 70.96                      & [70.73,71.20] \\ 
                             & \multirow{2}{*}{100} & $0.20\sigma$ & 68.49                      & [68.25,68.73] \\ 
                             &     & $0.25\sigma$ & 69.27                      & [69.04,69.49] \\ 
                             & \multirow{2}{*}{200} & $0.20\sigma$ & 61.28                      & [61.06,61.53] \\ 
                             &     & $0.25\sigma$ & 60.78                      & [60.53,61.02] \\
\bottomrule
\end{tabular}
\end{table*}

However, while our method provides a more precise variance estimate, it does not guarantee superior SampEn hyperparameter choices. This aspect is further explored in Section \ref{section:synthetic-param-selection}.

\subsubsection{Impact of Regularization on BO-driven SampEn Hyperparameter Search}
\label{subsubsection:picking-lambda}

In Eq. (\ref{eq:mse-optimization}), we introduced the regularization function $\Omega(r) = \lambda \sqrt{r}$ for $\lambda \geq 0$ to prevent trivial solutions where $r \longrightarrow \infty$. This regularization is essential because selecting $r$ greater than the maximum $L_\infty$ distance between two templates would yield zero MSE and a meaningless characterization of time series complexity.

Two primary considerations arise when applying regularization to the objective function. First, determining the appropriate magnitude of $\lambda$. Second, assessing whether the level of regularization applied is sufficient.

\begin{figure*}
    \centering
    \includegraphics[width=\linewidth]{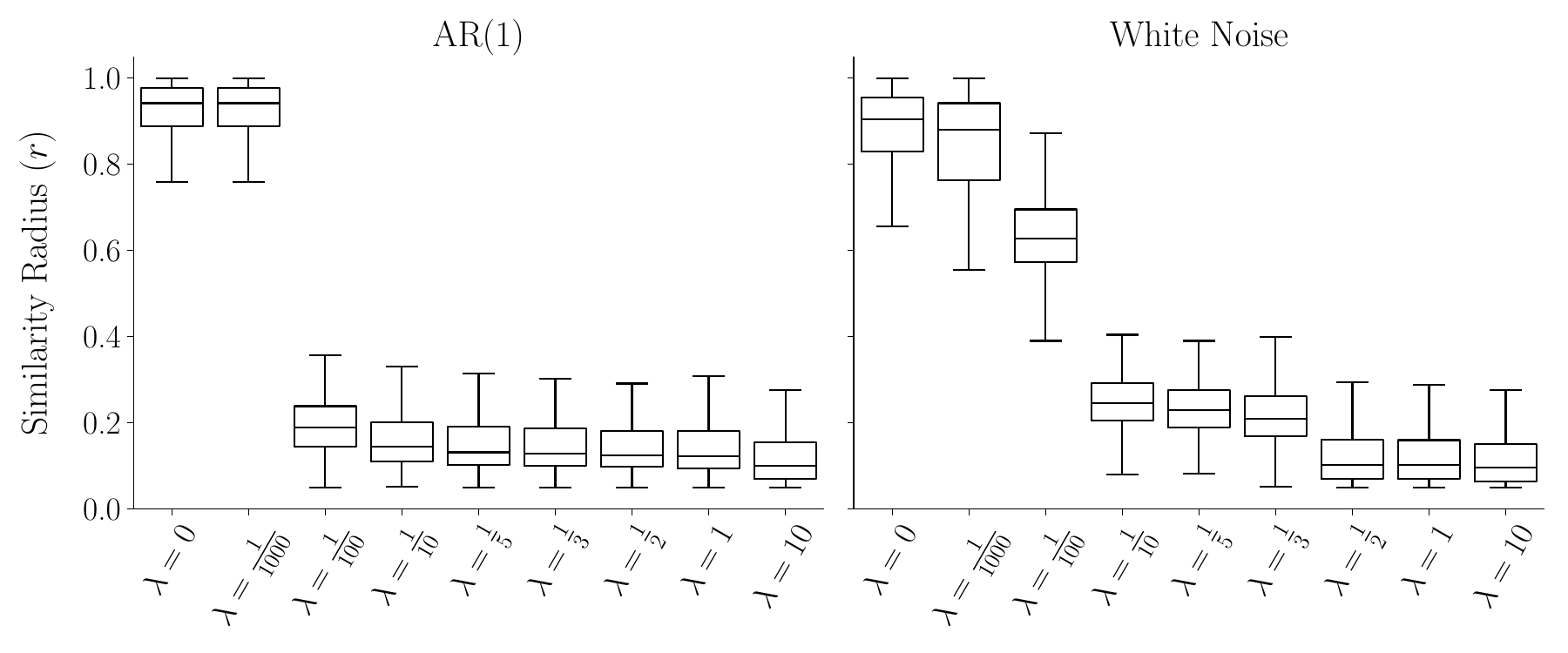}
     \caption{$\lambda \in \left[\frac{1}{10}, \frac{1}{3} \right]$ appears to effectively regularize the optimization procedure, preventing fixation at or near $r = 1$, while also avoiding fixation near the lower bound such as with $\lambda = 1$ or $\lambda = 10$.}
     \label{fig:regularization-effect}
\end{figure*}

To address these questions, we focused on selecting optimal SampEn hyperparameters $(m, r)$ for AR(1) and Gaussian white noise signals, $\boldsymbol{x} \in \mathbb{R}^{100}$, each containing $N = 100$ observations with fixed values for $\phi$ and $\sigma$. We employed Algorithm \ref{alg:sampen-bayes-opt}, setting $q = 0.5$ for AR(1) signals and $q = 0.9$ for white noise signals, with $B = 100$ bootstrap replicates and $\widetilde{T} = 100$ Bayesian optimization (BO) trials. We varied $\lambda$ across the range $\left\{0, \frac{1}{1000}, \frac{1}{100}, \frac{1}{10}, \frac{1}{5}, \frac{1}{3}, \frac{1}{2}, 1, 10 \right\}$. Fig. \ref{fig:regularization-effect} illustrates the distribution of $r$ values explored by the BO algorithm during optimization.

Overall, it appears that $\lambda \in \left[\frac{1}{10}, \frac{1}{3} \right]$ encourages the BO algorithm to avoid fixating on $r = 1$, as observed with $\lambda = 0$ and $\lambda = \frac{1}{1000}$, or fixating on the lower bound of $r$ with $\lambda = 1$ or $\lambda = 10$. Ideally, we aim to select a radius value that yields a search within the recommended range of $[0.10, 0.25] \times \sigma$ \citep{delgado2019approximate}, while allowing flexibility for achieving better overall SampEn estimate stability.

We offer the following heuristic for selecting $\lambda$: if the BO algorithm primarily explores at or near the upper bound of the radius domain, consider increasing $\lambda$ to encourage tighter radii. Conversely, if the algorithm primarily focuses at or near the lower bound of the radius domain, suggesting that the objective is overly influenced by the regularization term $\lambda \sqrt{r}$, consider relaxing $\lambda$ to promote broader exploration of radius options.

\subsubsection{BO Search Dynamics for SampEn Hyperparameter Selection}
We assessed the quality of the BO search over the SampEn hyperparameter space by evaluating three properties:

\begin{itemize}
    \item The ability of the algorithm to find good, locally optimal solutions within a limited number of trials \citep{frazier2018bayesian}.
    \item The presence of distinct ``exploration'' and ``exploitation'' phases during the search process \citep{watanabe2023tree}.
    \item The algorithm's capability to identify the interdependence between $(m, r)$ and make appropriate trade-offs.
\end{itemize}

\begin{figure*}
    \centering
    \includegraphics[width=\linewidth]{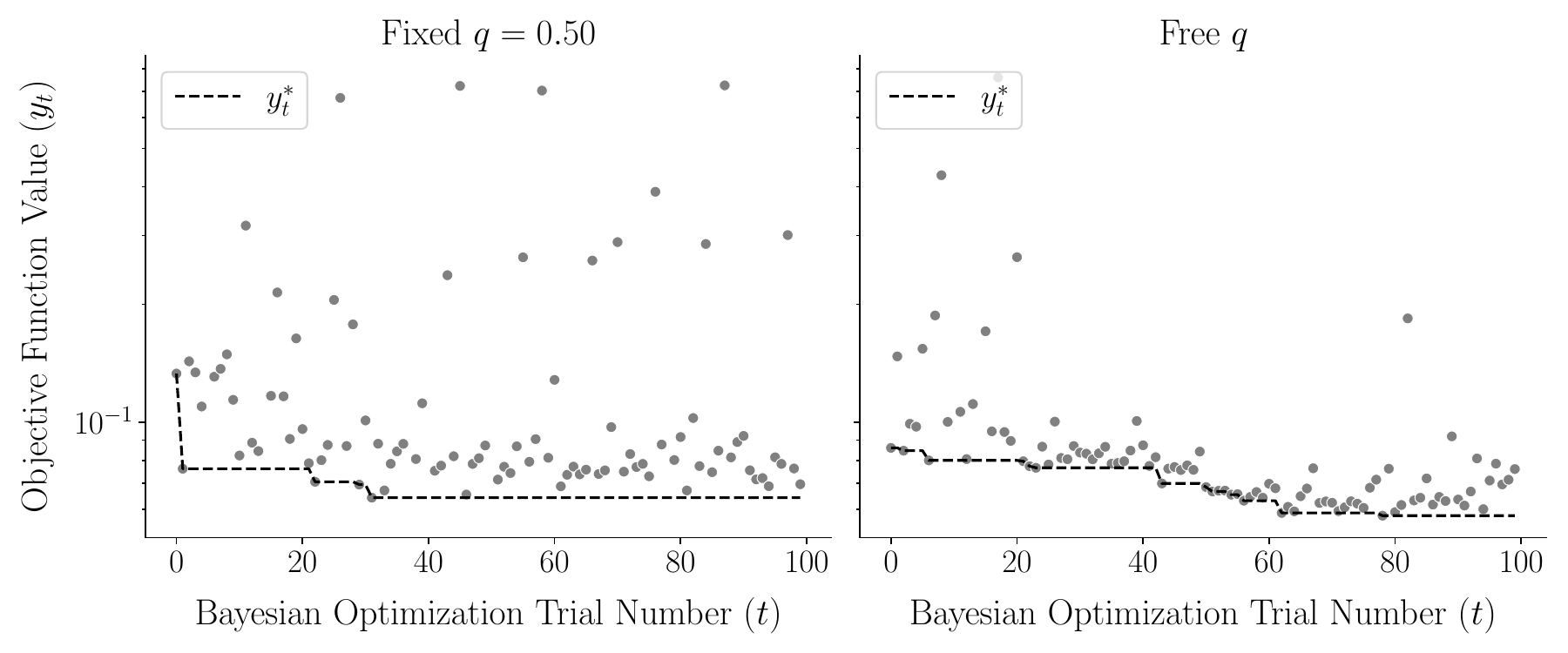}
    \caption{The BO algorithm identified good SampEn hyperparameter solutions in a relatively small number of trials for AR(1) signals, transitioning from exploration to exploitation. (Left): Without optimizing $q$, good solutions were found in fewer than 40 trials. (Right): Introducing $q$ as a decision variable increased complexity, but the algorithm still achieved a good solution in under 60 trials.}
    \label{fig:obj-over-time}
\end{figure*}

We investigated our proposed BO search process detailed in Algorithm \ref{alg:sampen-bayes-opt} using an AR(1) signal with $N = 100$ observations, $\phi = 0.9$, and $\sigma = 0.1$. With $\widetilde{T} = 100$ BO trials, $B = 100$ bootstrap replicates, and $\lambda = \frac{1}{10}$, we examined two scenarios: one with a fixed value, $q = 0.5$ and the other allowing the algorithm to determine $q^*$. The results are depicted in Fig. \ref{fig:obj-over-time}.

\begin{figure*}
    \centering
    \includegraphics[width=\linewidth]{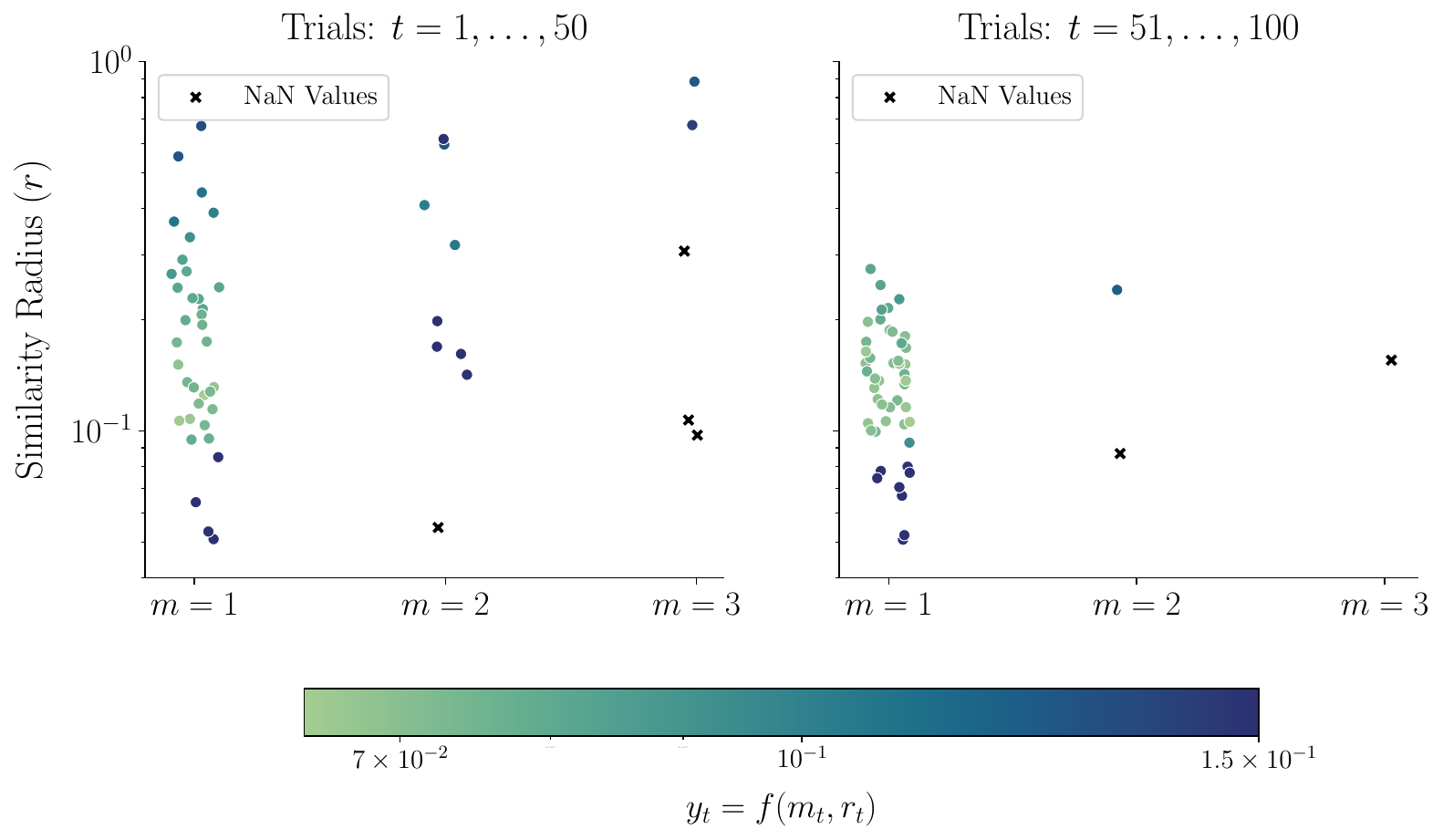}
    \caption{The BO algorithm adapts its search strategy from exploration to exploitation while accounting for the interdependence between $(m, r)$. (Left): In the initial 50 trials, the algorithm explored various $(m, r)$ combinations, adjusting the radius to obtain valid SampEn estimates. (Right): Subsequently, it focused on $m = 1$ to explore tighter similarity radii, effectively balancing the trade-off between larger $m$ values and smaller $r$ values.}
    \label{fig:bo-progress}
\end{figure*}

While introducing $q$ as a decision variable slightly increased the time to reach a good solution by approximately 20 trials, the BO search, in either case, transitioned to a high-performing regime after around 20 trials and shifted to an exploitation phase. Fig. \ref{fig:bo-progress}, focusing on $q = 0.5$, illustrates the combinations of $(m, r)$ explored by the algorithm over the computational trials.

The exploration and exploitation phases of the BO procedure are evident. Initially, the algorithm evaluates many $(m, r)$ combinations to understand the loss surface, whereas later, it primarily focuses on $m = 1$ while varying $r \in [0.10, 0.20]$. Moreover, the algorithm accounts for the interdependence between $(m, r)$. During exploration, it tests larger $m$ values with tighter radii, resulting in invalid SampEn estimates, and thus the radius to obtain valid estimates. Conversely, setting $m = 1$ enabled exploration of tighter radii, though extreme values of $r$ led to insufficient matches for a stable SampEn estimate. Thus, the algorithm balanced the desire for larger $m$ values and smaller $r$ values to achieve stable SampEn estimates.

\subsubsection{Sensitivity of SampEn Hyperparameter Selection to Bootstrap Success Probability}
\label{subsubsection:synthetic-q-effect}

Thus far, we have primarily used fixed values for the stationary bootstrap success probability hyperparameter, $q \in (0, 1)$. For white noise signals, we set $q = 0.9$ while for AR(1) signals, $q = 0.5$, based on heuristic assumptions regarding the overall temporal dependence structure of each signal class. However, given the introduction of an auxiliary variable into the optimization scheme, it becomes important to gauge the sensitivity of the optimization process to the choice of $q$.

\begin{figure*}
    \centering
    \includegraphics[width=\linewidth]{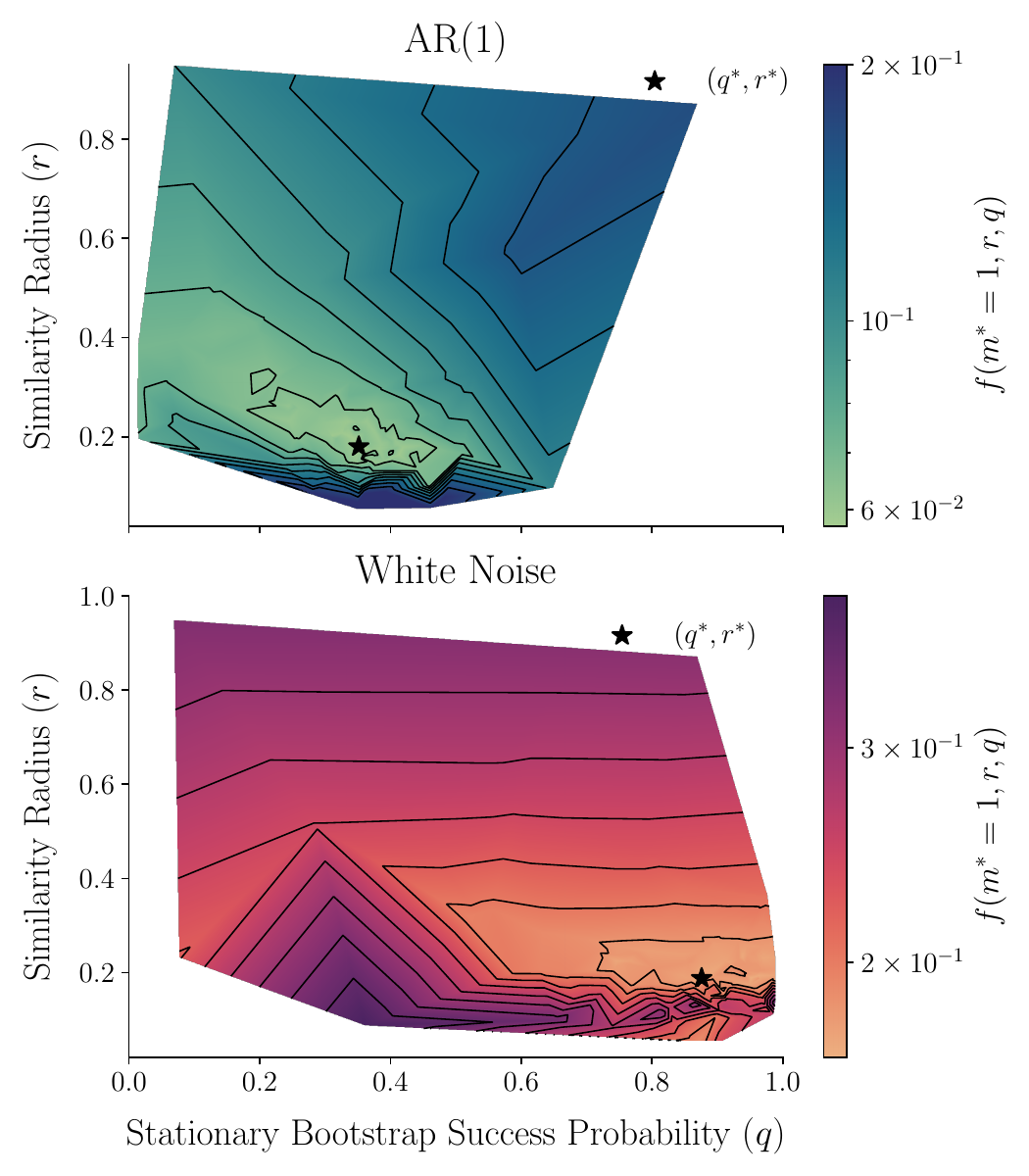}
    \caption{The BO algorithm was relatively insensitive to the block size success probability $q$ within a bounded range at the locally optimal similarity radius $r^*$. (Top): For AR(1) signals, $q$ values in the range $[0.20, 0.45]$ had approximately the same objective value at $r^*$, and even the heuristic suggestion of $q = 0.50$ did not have a notably greater loss value than than the optimal $q^*$. (Bottom): For white noise signals, $q$ values in the range $[0.75, 0.97]$ yielded nearly the same objective value at $r^*$.}
    \label{fig:q-vs-r-loss-surface}
\end{figure*}

To investigate this sensitivity, we examined the problem of selecting optimal values of $(r, q)$ for both white noise and AR(1) signals with $N = 100$. For AR(1) signals we set $\lambda = \frac{1}{10}$ and for white noise we set $\lambda = \frac{1}{3}$. Additionally we maintained $\widetilde{T} = 100$ BO trials, $B=100$ bootstrap replicates, and fixed $m = 1$, we constructed an interpolation of the $q$ versus $r$ loss surface for each signal class. Fig. \ref{fig:q-vs-r-loss-surface} illustrates the outcomes of this analysis.

Overall, we found that at the optimal $r^*$, the BO algorithm exhibits relatively low sensitivity to variations in $q$ within a bounded range. For white noise signals, $q \in [0.75, 0.97]$ yielded nearly identical objective values at $r^*$, while for AR(1) signals, $q \in [0.20, 0.45]$ showed comparable results. Notably, even when deviating from the optimal $q^*$, as observed in the AR(1) signal case with our heuristic suggestion of $q = 0.50$, the objective value remained relatively close. This suggests that the choice of $q$ may benefit from the stochastic nature of the signal bootstrapping scheme, enhancing the overall robustness of the optimization process \citep{politis2004automatic}.

\subsubsection{Benchmarking SampEn Hyperparameter Selection Methods}
\label{section:synthetic-param-selection}
In this section, we evaluate our method's performance in selecting SampEn hyperparameters relative to existing benchmarks. We compare our approach to the optimization-based method by \citet{lake2002sample}, the SampEn convergence-based approach, and the canonical SampEn hyperparameter settings of $(m = 2, r = 0.20\sigma)$.

We aim to select an optimal $(m, r)$ combination from a signal set, $\boldsymbol{S} = \left\{\boldsymbol{x}_1, \ldots, \boldsymbol{x}_n\right\}$, where $n = 100$ signals, each containing $N = 100$ observations. We explore this task for both white noise signals and AR(1) signals.

We assess the SampEn estimation error using the regularized MSE objective function (Eq. \ref{eq:modified-obj}) for each approach and measure the computation time required to reach the locally optimal $(m, r)$ combination. Since both the optimization-based approach by Lake et al. and the convergence-based approach do not inherently generate uncertainty bounds of the SampEn estimate, we approximate the mean regularized SampEn estimate MSE using a Gaussian distribution, as suggested by \citet{lake2002sample}. See Appendix \ref{appendix:sampen-gaussian-approx} for further details.

For the optimization-based approach, we search over a discretized grid of similarity radii values, $\mathcal{R}$, given a fixed embedding dimension value ($m = 1$). We search over $\mathcal{R} = \{0.10\sigma, 0.15\sigma, \ldots, 0.95\sigma, \sigma\}$ where $\sigma = 1$ and use a linear interpolation scheme to discretize to a grid with a step size of $0.01$.

For the convergence-based approach, we fix $m = 1$ and use the SampEn variance estimator proposed by \citet{lake2002sample} to calculate the median SampEn estimate variance across $\boldsymbol{S}$ for a given $(m = 1, r)$ combination. We select the radius value by identifying the ``elbow'' in the radius versus SampEn variance curve using the knee-finding algorithm by \citet{satopaa2011finding}.

In contrast, for our BO procedure for determining optimal SampEn hyperparameters, we fixed $\lambda = \frac{1}{10}$ for AR(1) signals and $\lambda = \frac{1}{3}$ for Gaussian white noise signals. We limited $m \in \{1, 2, 3\}$ and allotted $B = 100$ bootstrap samples with a computational budget of $\widetilde{T} = 100$ trials. We also compared the SampEn parameter selection strategies to the canonical $(m = 2, r = 0.20\sigma)$ setting, using our proposed BO procedure to select $q^*$ with the same $\lambda$, $B$, and $\widetilde{T}$ settings. Table \ref{tab:synthetic-opt-comparison} presents the results of this analysis.

\begin{table}[htbp]
\centering
\caption{Our proposed BO method demonstrates statistically significantly lower regularized mean SampEn estimate MSE compared to competing benchmarks for both signal classes ($p < 0.05$, one-sided Mann-Whitney U Test). Although methods by \protect\citet{lake2002sample} and the convergence-based approach entail lower computational costs than our BO routine, all methods achieve results within a reasonable timeframe. Notably, there seems to be little difference in computational time between our BO-based method searching for $(m, r, q)$ and the standard SampEn approach, which focuses solely on optimizing the stationary bootstrap success probability, $q$.}
\label{tab:synthetic-opt-comparison}
\resizebox{\linewidth}{!}{%
\begin{tabular}{cccccclc}
\toprule
\multicolumn{1}{c}{Signal Type} & Method          & MSE                         & $p$                     & Time (sec)         & $m^*$ & $r^*$             & $\text{SampEn}(m^*, r^*)$ \\
\midrule
\multirow{4}{*}{White Noise} & SampEnEff       & $0.282 \pm 0.006$           & $\leq 3.398 \times 10^{-8}$ & $0.116 \pm 0.006$  & 1     & $0.668 \pm 0.029$ & $1.021 \pm 0.042$         \\
                             & Convergence     & $0.191 \pm 0.008$           & $0.043$                     & $0.035 \pm 0.003$  & 1     & $0.253 \pm 0.034$ & $1.974 \pm 0.142$         \\
                             & Standard SampEn & $0.574 \pm 0.061$           & $3.139 \times 10^{-8}$      & $17.262 \pm 0.599$ & 2     & 0.2               & $2.301 \pm 0.048$         \\
                             & Ours            & $\mathbf{0.187 \pm 0.003}$  &                             & $21.671 \pm 0.354$ & 1     & $0.194 \pm 0.016$ & $2.232 \pm 0.079$         \\
\addlinespace
\multirow{4}{*}{AR(1)}       & SampEnEff       & $0.089 \pm 0.002$           & $\leq 3.398 \times 10^{-8}$ & $0.115 \pm 0.002$  & 1     & $0.775 \pm 0.038$ & $0.376 \pm 0.028$         \\
                             & Convergence     & $0.063 \pm 0.003$           & $\leq 3.398 \times 10^{-8}$ & $0.033 \pm 0.001$  & 1     & $0.308 \pm 0.054$ & $1.051 \pm 0.147$         \\
                             & Standard SampEn & $0.138 \pm 0.007$           & $\leq 3.398 \times 10^{-8}$ & $16.71\pm 0.20$    & 2     & 0.2               & $1.450 \pm 0.022$         \\
                             & Ours            & $\mathbf{0.056 \pm 0.0005}$ &                             & $17.064 \pm 0.371$ & 1     & $0.241 \pm 0.012$ & $1.255 \pm 0.049$         \\
\bottomrule
\end{tabular}
}
\end{table}

The results from Table \ref{tab:synthetic-opt-comparison} highlight several findings. Our BO approach consistently outperforms the comparable benchmarks in terms of the regularized MSE for all signal types. Additionally, the SampEnEff objective selected much larger radii than either our method or the SampEn convergence-based strategy, which may risk losing too much detailed system information \citep{pincus1994physiological} and may be attributed to the variance estimation issue we previously identified in Section \ref{section:variance-estimation}.

\subsubsection{Computational Cost of BO-driven SampEn Hyperparameter Selection}
\label{section:computational-efficiency}

\begin{figure*}[htbp]
    \centering
    \includegraphics[width=\linewidth]{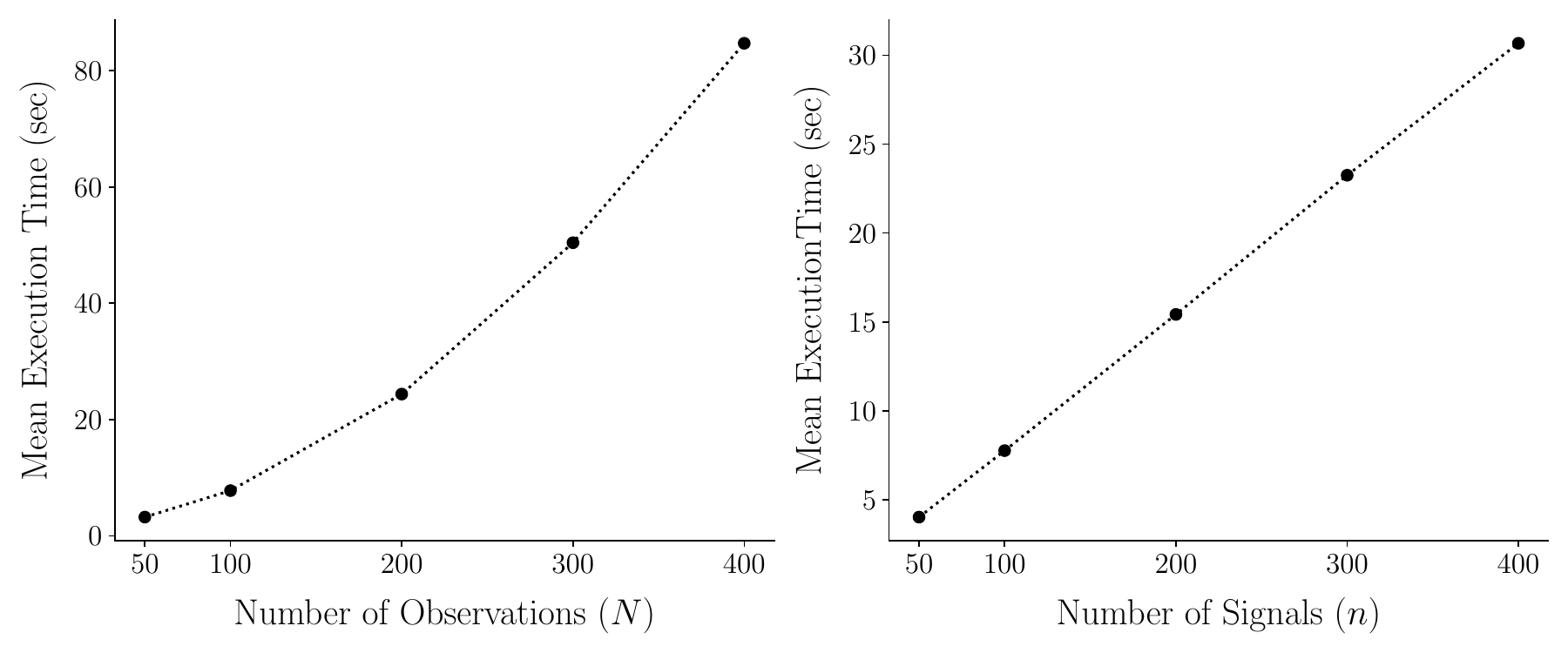}
    \caption{Computation execution time for our proposed SampEn optimization algorithm varies with signal length and number of signals. While execution time grows linearly with the number of signals for a fixed length, it grows non-linearly as the signal length increases, likely due to the $\mathcal{O}(N^2)$ computational complexity of the SampEn algorithm \protect\citep{jiang2011fast}.}
    \label{fig:execution-time}
\end{figure*}

Our BO-based algorithm outperforms existing benchmarks in terms of SampEn variance estimation and overall estimation error but comes with increased computational expense. To quantify this, we assessed the time required to reach an optimal, unified $(m, r, q)$ solution across a signal set $\boldsymbol{S}(N)$, varying signal length ($N$) and the number of signals ($n$), and repeated the experiment ten times. Using AR(1) signals with the same parameter settings for $\lambda$, $B$, and $\widetilde{T}$ as detailed in Section \ref{section:synthetic-param-selection}, we observed that computation time grows linearly with the number of signals but non-linearly with signal length likely due to the $\mathcal{O}(N^2)$ computational complexity of the SampEn algorithm \citep{jiang2011fast}. Fig. \ref{fig:execution-time} displays the results of this experiment.

While our approach yields practical results within reasonable computation times, it may not be suitable beyond signal lengths of approximately $N \approx 500$ observations. Beyond this threshold, standard SampEn parameters or simpler optimization methods may be more appropriate, considering SampEn's relative stability.

\subsection{Entropy Analysis on Short-Signal Benchmarks}
\label{subsection:real-experiments}

To conclude our experiments, we evaluated the performance of our SampEn hyperparameter selection optimization method against conventional benchmarks across five distinct short-signal datasets:

\begin{enumerate}
    \item Electrocardiogram (ECG) signals differentiating myocardial infarction from normal heartbeats \citep{olszewski2001generalized}, with $N = 96$ observations per signal.
    \item Traffic loop sensor data from near the Los Angeles Dodgers stadium, comparing traffic during games on weekends versus weekdays \citep{misc_dodgers_loop_sensor_157}, with $N = 288$ observations per signal.
    \item Accelerometer readings (roll) from a robot navigating over cement and carpet surfaces \citep{mueen2011logical}, with $N = 70$ observations per signal.
    \item Sensor data from silicone wafer production, distinguishing between normal and defective items \citep{olszewski2001generalized}, with $N = 152$ observations per signal.
    \item Sensor readings distinguishing between humidity signals versus temperature signals \citep{sun2005online}, with $N = 84$ observations per signal.
\end{enumerate}

Each dataset contains two classes with known ground truth. Our focus was on evaluating SampEn hyperparameter selection methods' ability to discern these known differences rather than employing a supervised machine learning approach for class segregation. This approach showcases the algorithm's potential efficacy in scenarios lacking known ground truths or where class discrimination is not the primary objective. Furthermore, by utilizing signals across diverse applications, including medical diagnostics, traffic analysis, and sensor-based monitoring, we highlight SampEn's versatility in data scientific applications.

We ensured signal stationarity, a necessary condition for valid entropy analysis \citep{chatain2019effects}, using the Augmented Dickey-Fuller test \citep{dickey1979distribution}, with the Holm-Sidak method \citep{holm1979simple, vsidak1967rectangular} applied to adjust for multiple testing errors at a $\alpha = 0.05$ confidence level. Detailed pre-processing steps are documented in Appendix \ref{appendix:weakly-stationary-signals}.

We compared several SampEn hyperparameter selection strategies, including our BO algorithm, the optimization-based criterion detailed by \citet{lake2002sample}, convergence-based methods, standard SampEn parameters $(m = 2, r = 0.20\sigma)$, and Fuzzy Entropy (FuzzEn) as an alternate measure \citep{chen2007characterization}. FuzzEn employs a fuzzy similarity condition, differing from SampEn’s hard similarity criteria, with standard parameters set to $(m = 2, r = 0.20\sigma, \eta = 2)$.

For the SampEnEff objective and the convergence-based approach, we utilized the same similarity radius grid of $\mathcal{R} = \{0.10\sigma, 0.15\sigma, \ldots, 0.95\sigma, \sigma\}$ using the linear interpolation scheme with a grid fidelity of $0.01$, and selected $m$ using the autoregressive heuristic proposed by \citet{lake2002sample}. Specifically, we found the optimal median time order lag using the Bayesian information criterion and selected that value as a proxy for embedding dimension \citep{lake2002sample, schwarz1978estimating}.

For our BO optimization approach, we set $B = 100$ bootstrap replicates and $\widetilde{T} = 200$ BO trials with $\lambda = \frac{1}{3}$ for all of the signal sets. 

For each signal set, we analyzed the locally optimal $(m^*, r^*)$ combination selected by each method, the median standard error (SE) of the SampEn estimates, and the Mann-Whitney U-test p-value comparing the distribution of SampEn estimates at $(m^*, r^*)$.

The outcome of this analysis is summarized in Table \ref{tab:benchmark-summary-results}. We discuss three main findings.

First, our BO framework consistently identified statistically significant differences between classes across all five signal sets at an $\alpha = 0.05$ significance level, unlike competing methods. Particularly, for ECG data, Dodger's stadium traffic, and robot surface accelerometer readings, our algorithm was the sole approach detecting known differences.

Second, SampEnEff generally under-performed, failing to establish statistically significant distinctions between the signal groups. This could be attributed to its preference for larger similarity radii due to previously discussed issues of SampEn variance estimation issues, especially in shorter signal sets. Conversely, the convergence-based method outperformed SampEnEff in all datasets except the ECG signal set.

Third, standard SampEn parameters and FuzzEn detected significant differences in specific datasets, like wafer manufacturing and humidity versus temperature signal sets. However, predicting the effectiveness of these parameters without prior analysis is challenging. Moreover, these conventional approaches lack the capability to quantify the uncertainty associated with entropy estimates. Although fixing parameters, such as $m = 2$ and $r=0.20\sigma$, and finding $q^*$ to estimate uncertainty is feasible, our analysis demonstrated a relatively small computational difference to optimize $(m^*, r^*, q^*)$ compared to merely $q^*$.

\begin{table*}
\centering
\caption{Our proposed BO method was the only approach to consistently detect known, statistically significant differences at $\alpha = 0.05$ in entropy between signal classes across all benchmark datasets while achieving comparable or lower SampEn estimate standard error. Results of the entropy analysis are summarized below, including the the Mann-Whitney U-test p-value $(p)$, and the median SampEn standard error, $\text{Median } \text{SE}(\hat{\theta}(m^*, r^*))$, at the optimal parameters $(m^*, r^*)$ selected by each method.}
\label{tab:benchmark-summary-results}
\begin{tabular}{cccccc}
\toprule
Data & Method & $m^*$ & $r^*$ & Median $\text{SE}\left(\hat{\theta}(m^*, r^*) \right)$ & $p$ \\
\midrule
\multirow{5}{*}{\shortstack{ECG \citep{olszewski2001generalized}\\$n = 56$\\$N = 96$}} 
& Ours            & 1     & 0.11  & 0.19 & \textbf{0.042} \\
& SampEnEff       & 4     & 0.50  & 0.19 & 0.35 \\
& Convergence     & 4     & 0.41  & 0.23 & 0.47 \\
& Standard SampEn & 2     & 0.20  &      & 0.21 \\
& FuzzEn          & 2     & 0.20  &      & 0.30 \\
\addlinespace
\multirow{5}{*}{\shortstack{Dodgers' Traffic \citep{misc_dodgers_loop_sensor_157}\\$n = 144$\\$N = 288$}} 
& Ours            & 1     & 0.18  & 0.097 & \textbf{0.002} \\
& SampEnEff       & 3     & 0.40  & 0.28  & 0.79 \\
& Convergence     & 3     & 0.34  & 0.29  & 0.32 \\
& Standard SampEn & 2     & 0.20  &       & 0.30 \\
& FuzzEn          & 2     & 0.20  &       & 0.73 \\
\addlinespace
\multirow{5}{*}{\shortstack{Robot Surface \citep{mueen2011logical}\\$n = 511$\\$N = 70$}}
& Ours            & 1     & 0.05  & 0.18 & $\mathbf{2.21 \times 10^{-5}}$ \\
& SampEnEff       & 2     & 0.55  & 0.19 & 0.73 \\
& Convergence     & 2     & 0.10  & 0.36 & 0.12 \\
& Standard SampEn & 2     & 0.20  &      & 0.12 \\
& FuzzEn          & 2     & 0.20  &      & 0.67 \\
\addlinespace
\multirow{5}{*}{\shortstack{Wafer Manufacturing \citep{olszewski2001generalized}\\$n = 1000$\\$N = 152$}}
& Ours            & 1     & 0.10  & 0.022 & $\mathbf{5.11 \times 10^{-5}}$ \\
& SampEnEff       & 1     & 1.00  & 0.021 & 0.29 \\
& Convergence     & 1     & 0.24  & 0.023 & $\mathbf{1.62 \times 10^{-6}}$ \\
& Standard SampEn & 2     & 0.20  &       & 0.16 \\
& FuzzEn          & 2     & 0.20  &       & $\mathbf{8.40 \times 10^{-5}}$ \\
\addlinespace
\multirow{5}{*}{\shortstack{Sensor Modality \citep{sun2005online}\\$n = 825$\\$N = 84$}}
& Ours            & 1     & 0.10  & 0.19  & \textbf{0.006} \\
& SampEnEff       & 1     & 1.00  & 0.056 & 0.56 \\
& Convergence     & 1     & 0.48  & 0.12  & 0.45 \\
& Standard SampEn & 2     & 0.20  &       & $\mathbf{2.61 \times 10^{-3}}$ \\
& FuzzEn          & 2     & 0.20  &       & 0.39 \\
\bottomrule
\end{tabular}
\end{table*}

\section{Discussion and Conclusion}
\label{section:discussion}

In this study, we evaluated our proposed SampEn hyperparameter optimization approach against existing optimization-based, convergence-based, and standard SampEn parameter selection strategies using synthetic signals and publicly available benchmarks.

For the synthetic experiments, our main findings are as follows:

\begin{itemize}
    \item Our bootstrap-based SampEn variance estimator consistently achieved 60 to 90 percent lower estimation error compared to Lake et al.'s estimator across multiple signal types, similarity radii, and signal lengths.
    \item A regularization value of $\lambda \in \left[\frac{1}{10}, \frac{1}{3}\right]$ effectively regulated the BO algorithm, preventing fixation on extreme bounds of the similarity radius domain.
    \item The BO algorithm typically converged to satisfactory solutions in under 75 computational trials and demonstrated a balanced exploration-exploitation approach within the $(m, r, q)$ decision space.
    \item Sensitivity analysis revealed the BO algorithm's robustness to the specification of $q$ at $r^*$.
    \item Our proposed algorithm yielded a statistically significantly lower mean regularized SampEn estimate MSE, with competing methods often favoring larger similarity radii.
    \item Our method found locally optimal, unified $(m, r, q)$ combinations across signal sets of varying signal lengths and signal set sizes in a reasonable timeframe. However, for signals longer than approximately $N \approx 500$ observations, other methods may be more appropriate.
\end{itemize}

Across five diverse, short-signal benchmarks, our method was the only one to identify known, statistically significant differences between signal classes. This appears to be driven by our joint consideration of $(m, r)$ selection for stable SampEn estimation and preference for tighter similarity radii.

Our results suggest that the key advantages of our approach are its regularization against overly large radii and more accurate quantification of estimation uncertainty, leading to superior performance on short-signal scenarios.

There are, however, open questions regarding entropy analysis with very short signals ($N \leq 50$), where SampEn's reliability diminishes, sometimes yielding undefined values due to lack of template matches \citep{li2015assessing}. This limitation is particularly relevant for applications like cardiopulmonary exercise testing, where signals may contain fewer than 50 observations \citep{takken2009cardiopulmonary}.

To address this challenge, future work could explore two main directions: improving existing entropy measures or modifying the SampEn computation itself. Permutation entropy (PermEn) shows promise for shorter signals \citep{bandt2002permutation, cuesta2019embedded}, with recent research demonstrating reliable PermEn estimates using a Bayesian framework with as few as $N = 10$ observations \citep{blankspermen2024}. Alternatively, modifying the SampEn computation, as suggested by \citet{richman2004sample} to ensure template matches for shorter signals may be beneficial.

Ultimately, addressing the challenges of entropy estimation for very short signals is necessary for expanding the applicability of these complexity measures to a wider range of real-world, data-scientific scenarios.

\bibliographystyle{unsrtnat}

\appendix

\section{Tree-structured Parzen Estimator Hyperparameters}
\label{appendix:tpe-hyperparams}

In the Tree-structured Parzen Estimator (TPE) optimization framework, there are four hyperparameters of note:

\begin{itemize}
    \item The top quantile value, $\gamma \in (0, 1]$ which delineates the better-performing observation set, $\mathcal{D}^{(l)}$, and worse-performing observation set, $\mathcal{D}^{(g)}$.
    \item The kernel model: $k_d: \Psi_d \times \Psi_d \rightarrow \mathbb{R}_+$ describes the distribution of the $d$-th decision variable given the observation set.
    \item The bandwidth parameters, $b^{(l)}, b^{(g)} \in \mathbb{R}_+$ control the kernel estimators for the better and worse-performing observation sets
    \item $\left\{w_t \right\}_{t=0}^{T+1}$ assign weights to the Gaussian mixture components.
\end{itemize}

These hyperparameters use default values from the ``Optuna'' framework \citep{optuna_2019} for the TPE optimizer. The following sections provide detailed descriptions of each component as applied to the SampEn hyperparameter selection problem.

\subsection{Top Quantile Selection}
\label{subsection:top-quantile-selection}
The top quantile value, $\gamma \in (0, 1]$, controls the sizes of the sets $\mathcal{D}^{(l)}$ and $\mathcal{D}^{(g)})$. For all iterations, we use $\gamma = \Gamma(T) = 0.10$. However, when selecting the $(T+1)$-th point, the size of the better-performing set $\mathcal{D}^{(l)}$ is adjusted according to:

\begin{equation}
    T^{(l)} = \min\left(\lceil \gamma T \rceil, 25 \right).
    \label{eq:better-set-size}
\end{equation}

Given a sorted observation set, $\mathcal{D} = \left\{(\boldsymbol{\psi}_t, y_t) \right\}_{t=1}^T$, on $y_t$, such that $y_1 \leq y_2 \leq \cdots \leq y_T$, we define $\mathcal{D}^{(l)} = \left\{(\boldsymbol{\psi}_t, y_t) \right\}_{t=1}^{T^{(l)}}$ and $\mathcal{D}^{(g)}$ contains the remaining objective function evaluations.

\subsection{Gaussian Kernel Functions and Bandwidth Selection}
\label{subsection:kernel-and-bandwidth}
To construct a mixture of Gaussian kernels, we define kernel functions for our SampEn hyperparameters: $(m, r, q)$. These kernels, denoted as $k_d: \Psi_d \times \Psi_d \rightarrow \mathbb{R}$, are defined over the domains of the respective decision variables. Given that $m \in \Psi_1 \subseteq \mathbb{Z}_+$, and $r, q \in \Psi_2, \Psi_3 \subseteq (0, 1)$, we choose appropriate kernels and bandwidths for each.

For the continuous variables $r$ and $q$, we employ the canonical Gaussian kernel, parameterized by $b \in \mathbb{R}_+$. The kernel for the $t$-th observation of $r$ (similarly for $q$) is given by:

\begin{equation}
    g(r, r_t \mid b) = \frac{1}{\sqrt{2\pi b^2}}\exp\left(-\frac{1}{2} \left(\frac{r - r_t}{b} \right)^2 \right).
\end{equation}

However, since $r$ and $q$ are bounded within the unit interval, we utilize a truncated Gaussian kernel:

\begin{equation}
    k_d(r, r_t) = \frac{1}{\int_0^1 g(r^\prime, r_t \mid b) \ dr^\prime} \cdot g(r, r_t \mid b). 
    \label{eq:radius-gaussian-kernel}
\end{equation}

For the embedding dimension $m$, a discrete parameter, we adopt a similar approach but with slight modifications due to its discrete nature. We limit $m$ to a finite set ${1, 2, \ldots, U}$, where $U$ is the upper bound of possible $m$ values. The kernel for the $t$-th observation of $m$, given the bandwidth parameter $b$, is computed as:

\begin{equation}
    k_1(m, m_t) = \frac{1}{\int_{1 - 1/2}^{U + 1/2} g(m^\prime, m_t \mid b) \ dm^\prime} \times
    \int_{1-1/2}^{U+1/2} g(m^\prime, m_t \mid b) \ dm^\prime.
\end{equation}

We select the bandwidth parameter $b$ using Scott's rule\citep{scott2015multivariate} with the ``magic-clipping'' heuristic \citep{bergstra2011algorithms}. Specifically, $b$ is calculated as:

\begin{equation}
    b = T^{-\frac{1}{d + 4}},
\end{equation}

where $d$ is the dimensionality of the kernel. We also introduce $b_{\min}$:

\begin{equation}
    b_{\min} = \frac{U -L}{\min\left(\{T, 100 \} \right)},
\end{equation}

where $U$ and $L$ indicate the upper and lower bounds of the decision variable. Finally, we set the bandwidth parameter as:

\begin{equation}
    b^\prime = \max\left(\{b, b_{\min} \} \right).
    \label{eq:final-bandwidth}
\end{equation}

\subsection{Gaussian Kernel Mixture Weights}
\label{subsection:gaussian-weights}
In our Bayesian optimization framework, we utilize a mixture of Gaussian kernels for the surrogate model. This requires a set of weights, $\left\{w_t \in [0, 1] \right\}_{t=0}^{T+1}$, which we define using the ``old decay'' weighting scheme proposed by \citet{bergstra2013making} defined as:

\begin{equation}
    w_t = \begin{cases}
        \frac{1}{T^{(l)} + 1}, &\text{if, } t = 0, \ldots, T^{(l)} \\
        \frac{w_t^\prime}{\sum_{k=T^{(l)} + 1}^{T+1} w_k^\prime}, &\text{if, } t = T^{(l)} + 1, \ldots, T+1
    \end{cases}
    \label{eq:weighting-scheme}
\end{equation}

In Eq. (\ref{eq:weighting-scheme}), for $t = T^{(l)} + 1, \ldots, T+1$, we compute $w_t^\prime$ according to:

\begin{equation}
    w_t^\prime = \begin{cases}
    1, &\text{if } i_t > T^{(g)} + 1 - 25 \\
    \tau(i_t) + \frac{1 - \tau(i_t)}{T^{(g)} + 1}, &\text{else}
    \end{cases},
\end{equation}

where $i_t$ represents the query order, with $i_t = 1$ being the oldest query and $i_t = T^{(g)} + 1$ the most recent one. The decay rate $\tau(i_t)$ is defined as $(i_t - 1)/(T^{(g)} - 25)$ \citep{bergstra2013making}. This weighting scheme assigns less weight to older queries in $\mathcal{D}^{(g)}$ and greater weight to more recent ones.

\subsection{Constructing the Surrogate Model}
\label{subsection:surrogate-model}
The surrogate models $p\left(\boldsymbol{\psi} \mid \mathcal{D}^{(l)} \right)$ and $p\left(\boldsymbol{\psi} \mid \mathcal{D}^{(g)} \right)$ are built upon a non-informative Gaussian prior, $p_0$. Considering an upper bound $U$ for the embedding dimension $m$, we define the mean vector $\boldsymbol{\mu}_0$ and covariance matrix $\boldsymbol{\Sigma}_0$ as:

\begin{equation*}
    \boldsymbol{\mu}_0 = \left(\frac{U-1}{2}, \frac{1}{2}, \frac{1}{2} \right), \quad \boldsymbol{\Sigma}_0 = \begin{pmatrix}
        (U-1)^2 & 0 & 0 \\
        0 & 1 & 0 \\
        0 & 0 & 1 \\
    \end{pmatrix}.
\end{equation*}

The non-informative prior is then $p_0 \sim \mathcal{N}\left(\boldsymbol{\mu}_0, \boldsymbol{\Sigma}_0 \right)$. The surrogate model for the better-performing set is expressed as:

\begin{equation}
    p\left(\boldsymbol{\psi} \mid \mathcal{D}^{(l)} \right) = p_0(\boldsymbol{\psi}) + \prod_{d=1}^3 \sum_{t = 1}^{T^{(l)}} w_t \cdot k_d\left(\psi_d, \psi_{d, t} \mid b^{(l)} \right).
\end{equation}

Similarly, for the worse-performing group, we have:

\begin{equation}
    p\left(\boldsymbol{\psi} \mid \mathcal{D}^{(g)} \right) = p_0(\boldsymbol{\psi}) + \prod_{d=1}^3 \sum_{t=T^{(l)} + 1}^{T} w_t \cdot k_d\left(\psi_d, \psi_{d, t} \mid b^{(g)} \right),
\end{equation}

where $\psi_{d, t} \in \Psi_d$ represents the $d$-th dimension of the $t$-th observation, and $b^{(l)}$ and $b^{(g)}$ are the bandwidth parameters obtained using Eq. (\ref{eq:final-bandwidth}).

\section{Signal Set SampEn Hyperparameter Selection Optimization Algorithm}
\label{appendix:signal-set-opt-algo}

We provide a full specification of how to select locally optimal SampEn hyperparameters, $(m, r)$ for a signal set, $\boldsymbol{S} = \left\{\boldsymbol{x}_1, \ldots, \boldsymbol{x}_n \right\}$. This is detailed in Algorithm \ref{alg:sampen-bayes-opt-multiple-signals}.

\begin{algorithm*}
\caption{Signal Set SampEn Hyperparameter Selection Optimization}
\begin{algorithmic}[1]
\Require $\boldsymbol{S} = \left\{\boldsymbol{x}_1, \ldots, \boldsymbol{x}_n \right\}$, $B \in \mathbb{Z}_+$, $\widetilde{T} =$ Number of BO trials, $\lambda \geq 0$, $T_{\text{init}} \in \mathbb{Z}_+$
\State $\mathcal{D} \gets \emptyset$ \Comment{Empty observation set}
\For{$t = 1, \ldots, T_{\text{init}}$}
    \State Select $\boldsymbol{\psi}_t$ randomly
    \For{$i = 1, \ldots, n$}
        \State Compute the SampEn estimate of $\boldsymbol{x}_i$, $\hat{\theta}_i(m_t, r_t)$
        \State Generate bootstrap replicates of $\boldsymbol{x}_i$, $\left\{\boldsymbol{x}_{i, b} \right\}_{b=1}^B$, given $q_t$ \Comment{See Algorithm \ref{alg:stationary-bootstrap}}
        \State Compute bootstrap SampEn estimates, $\left\{\hat{\theta}_{i, b}(m_t, r_t) \right\}_{b=1}^B$
        \State $y_i = \text{MSE}\left(\hat{\theta}_{i}(m_t, r_t) \right) + \lambda\sqrt{r_t}$
    \EndFor
    \State $y_t = \frac{1}{n}\sum_{i=1}^n y_i$ \Comment{Mean objective value across the $n$ signals}
    \State $\mathcal{D} \gets \mathcal{D} \cup \left(\boldsymbol{\psi}_t, y_t \right)$
\EndFor
\While{$t \leq \widetilde{T}$}
    \State Select $\boldsymbol{\psi}_{t+1}$ using the TPE acquisition process \Comment{See Algorithm \ref{alg:tpe-new-point}}
    \For{$i = 1, \ldots, n$}
        \State Compute SampEn estimate of $\boldsymbol{x}_i$, $\hat{\theta}_i(m_{t+1}, r_{t+1})$
        \State Generate bootstrap replicates of $\boldsymbol{x}_i$, $\left\{\boldsymbol{x}_{i, b} \right\}_{b=1}^B$, given $q_{t+1}$ \Comment{See Algorithm \ref{alg:stationary-bootstrap}} 
        \State Compute bootstrap SampEn estimates, $\left\{\hat{\theta}_{i, b}(m_{t+1}, r_{t+1}) \right\}_{b=1}^B$
        \State $y_i = \text{MSE}\left(\hat{\theta}_{i}(m_{t+1}, r_{t+1}) \right) + \lambda\sqrt{r_{t+1}}$
    \EndFor
    \State $y_{t+1} = \frac{1}{n}\sum_{i=1}^n y_i$ \Comment{Mean objective value across the $n$ signals}
    \State $\mathcal{D} \gets \mathcal{D} \cup \left(\boldsymbol{\psi}_{t+1}, y_{t+1} \right)$
\EndWhile
\State $y^* \gets \min \ \left\{y_1, \ldots, y_{\widetilde{T}} \right\}$
\State $\boldsymbol{\psi}^* \gets \text{argmin} \ \left\{y_1, \ldots, y_{\widetilde{T}} \right\}$
\Return $\boldsymbol{\psi}^*, y^*$
\end{algorithmic}
\label{alg:sampen-bayes-opt-multiple-signals}
\end{algorithm*}

\section{SampEn Variance Estimators Comparison Experiment Details}
\label{appendix:sampen-variance-experiment-details}

This appendix details the methodology employed to compare the SampEn variance estimation between two proposed approaches. We begin by describing the normalized set of signals for each signal class (white noise and AR(1) signals), denoted as $\boldsymbol{S}(N) = \left\{\boldsymbol{x}_1, \ldots, \boldsymbol{x}_n \right\}$, where $n$ represents the number of generated signals, all with $N$ observations. We fix $n = 10,000$ and vary $N$ across ${50, 100, 200}$ to represent instances of short time series signals. Additionally, we evaluate $r \in \left\{0.20\sigma, 0.25\sigma \right\}$, where $\sigma$ is the standard deviation of the signal. All signals are normalized to have zero mean and unit variance.

We denote $\hat{\theta}_i(m, r)$ as the estimated SampEn of signal $\boldsymbol{x}_i$ at $(m, r)$, and $\left\{\hat{\theta}i(m,r) \right\}_{i=1}^n$ as the set of $n$ SampEn estimates at $(m, r)$ from the signal set, $\boldsymbol{S}(N)$. We approximate the ``true'' SampEn estimate variance at $(m, r, N)$ by the expression:

\begin{align}
\begin{aligned}
    \sigma^2(m, r, N) &= \mathbb{V}_N\left(\left\{\hat{\theta}_i(m,r) \right\}_{i=1}^n\right)\\
    &\approx \frac{1}{n-1} \sum_{i=1}^n \left(\hat{\theta}_i(m, r) - \bar{\theta}(m, r) \right)^2
    \label{eq:true-sampen-variance}
\end{aligned}
\end{align}

where $\bar{\theta}(m, r)$ is the mean of the SampEn estimate set.

We are interested in understanding how accurately a particular SampEn variance estimator performs. Given a signal, $\boldsymbol{x}_i$, for a particular estimator, $e$, we obtain a SampEn variance estimate: $\widehat{\sigma}_i^2(m, r, N, e)$. The mean squared error (MSE) of the SampEn variance estimator, $e$, for a particular signal class at $(m, r, N)$ is approximated as:

\begin{align}
\begin{aligned}
    \epsilon(m, r, N, e) &= \text{MSE}\left(\left\{\widehat{\sigma}_i(m, r, N) \right\}_{i=1}^{\widetilde{n}} \right)\\ 
    &\approx \frac{1}{\widetilde{n}} \sum_{i=1}^{\widetilde{n}} \left(\sigma^2(m, r, n) - \widehat{\sigma}_i^2(m, r, N, e) \right).
    \label{eq:sampen-variance-mse}
\end{aligned}
\end{align}

For a given signal type at $m = 1$, and fixed values for $(r, N)$, we generated 20 estimates of $\epsilon(m, r, N, e)$ for both our proposed bootstrap-based estimator and the estimator proposed by \citet{lake2002sample}. This process is summarized in Algorithm \ref{alg:variance-error-algo}.

\begin{algorithm*}
\caption{SampEn Variance Estimation Error Calculation Procedure}
\begin{algorithmic}[1]
\Require $N \in \mathbb{Z}_+$, $r \in \mathbb{R}_+$, Fix $m =1$, $q = 0.5$ (AR(1)) or $q = 0.9$ (White Noise), $B = 100$, $n = 10,000$, $\widetilde{n} = 100$
\State Generate $\boldsymbol{S}(N) = \left\{\boldsymbol{x}_1, \ldots, \boldsymbol{x}_n \right\}$ \Comment{Either a Gaussian white noise or AR(1) signal}
\State Calculate SampEn estimates from $\boldsymbol{S}(N)$: $\left\{\hat{\theta}_i(m, r) \right\}_{i=1}^n$
\State Compute $\sigma^2(m, r, N)$ using Eq. (\ref{eq:true-sampen-variance})
\For{$i = 1, \ldots, 20$}
    \State $\widetilde{\boldsymbol{S}}(N) \gets $ Randomly select $\widetilde{n}$ signals from $\boldsymbol{S}(N)$.
    \State Calculate $\left\{\widehat{\sigma}_j(m, r, N, \text{Counting}) \right\}_{j=1}^{\widetilde{n}}$ from the signals in $\widetilde{\boldsymbol{S}}(N)$ \Comment{See \citep{lake2002sample} for computation details}
    \State Calculate $\left\{\widehat{\sigma}_j(m, r, N, \text{Bootstrap}) \right\}_{j=1}^{\widetilde{n}}$ from the signals in $\widetilde{\boldsymbol{S}}(N)$ given $B$ bootstrap replicates and $q$ \Comment{See Section \ref{subsection:bootstrap-variance} and (Eq. \ref{eq:bootstrap-variance}) for details.}
    \State Calculate $\epsilon_i(m, r, N, \text{Counting})$ using Eq. (\ref{eq:sampen-variance-mse})
    \State Calculate $\epsilon_i(m, r, N, \text{Bootstrap})$ using Eq. (\ref{eq:sampen-variance-mse})
\EndFor
\Return $\left\{\epsilon_i(m, r, N, \text{Counting}) \right\}_{i=1}^{20}$ and $\left\{\epsilon_i(m, r, N, \text{Bootstrap}) \right\}_{i=1}^{20}$
\end{algorithmic}
\label{alg:variance-error-algo}
\end{algorithm*}

In Algorithm \ref{alg:variance-error-algo}, ``Counting'' refers to the SampEn variance estimation approach detailed by \citet{lake2002sample} and ``Bootstrap'' corresponds to our proposed estimation procedure. We repeated this process for $N \in {50, 100, 200}$ observations and $r \in {0.15\sigma, 0.20\sigma, 0.25\sigma }$.

We model the distribution of signal type $t$ of SampEn variance estimator errors using the following setup:

\begin{align}
\begin{aligned}
    \mu_{t, r, N, e} &\sim \mathcal{N}(0, 4) \\
    \sigma_{t, r, N, e} &\sim \mathcal{U}(0.01, 2.0) \\
    \epsilon_i(t, r, N, e) &\sim \log \mathcal{N}(\mu_{t, r, N, e}, \sigma_{t, r, N, e}) \\
    \Delta \epsilon(r, N) &\equiv \left(\frac{\exp(\mu_{t, r, N, 1}) - \exp(\mu_{t, r, N, 2})}{\exp(\mu_{t, r, N, 1})}\right) \times 100.
    \label{eq:variance-comparison-model}
\end{aligned}
\end{align}

In Eq. (\ref{eq:variance-comparison-model}), $\mu_{t, r, N, e}$ and $\sigma_{t, r, N, e}$ represent the mean and standard deviation of log-transformed SampEn variance MSE values, respectively.

We are particularly interested in the posterior distribution of $\Delta \epsilon(r, N)$, representing the mean percent difference in SampEn variance MSE between our bootstrap-based estimator ($\mu_{t, r, N, 2}$) and the counting-based approach ($\mu_{t, r, N, 1}$) detailed by \citet{lake2002sample}. We use the $\exp(\cdot)$ function to transform the values out of the log-error domain. Positive values of $\Delta \epsilon(m, r, N)$ would indicate that our proposed estimator has relatively lower SampEn variance estimator error.

We propose to approximate the posterior distribution of $\mu_{t, r, N, e}$ using the Markov Chain Monte Carlo (MCMC) algorithm via the No U-Turn Sampler algorithm in the PyMC probabilistic programming framework \citep{abril2023pymc, hoffman2014no}. To ensure MCMC convergence, we used trace plots, evaluated the effective sample size (ESS), and checked the Gelman-Rubin ``R-hat'' statistic \citep{gelman1992inference}. Trace plots visually assessed the chains' behavior, ensuring stability, good mixing, and a lack of trends, signifying successful exploration of the parameter space and convergence. ESS measurements quantified the number of independent samples, indicating efficient sampling and exploration of the posterior distribution. R-hat values close to one suggested convergence.

\section{Gaussian Approximation of SampEn Estimate Uncertainty}
\label{appendix:sampen-gaussian-approx}

Let $\boldsymbol{S} = \left\{\boldsymbol{x}_1, \ldots, \boldsymbol{x}_n\right\}$ be a signal set containing $n$ signals each with $N$ observations, and let $\left\{\hat{\theta}_i(m, r)\right\}_{i=1}^n$ be the set of SampEn estimates given a fixed $(m, r)$ obtained from $\boldsymbol{S}$. We are interested in generating an approximation of the mean squared error (MSE) of the set of SampEn estimates to compare alternative SampEn hyperparameter selection approaches against our proposed method. Approximating the SampEn MSE requires a way to reason about the uncertainty of the estimate.

\citet{lake2002sample} assert that provided $m$ is small enough and $r$ large enough to ensure a sufficient number of matches, SampEn can be assumed to be normally distributed. Specifically, the standard error (SE) of the SampEn estimate of $\boldsymbol{x}_i$ is approximated as:

\begin{equation}
    s_i \approx \frac{\sigma_{\text{CP}}}{\text{CP}},
    \label{eq:sampen-se-approx}
\end{equation}

where $\text{CP}$ is the conditional probability a signal will remain within $r$ for $m+1$ points given it has remained within $r$ for $m$ points, expressed as $\text{CP} = \frac{A^m(r)}{B^m(r)}$ (see \eqref{eq:sampen-bmr} and \eqref{eq:sampen-amr} for details on calculating $B^m(r)$ and $A^m(r)$, respectively). \citet{lake2002sample} provide a mechanism to compute $\sigma_{\text{CP}}$.

Under the Gaussian assumption, the SampEn of $\boldsymbol{x}_i$ at $(m, r)$ is then $\theta_i(m, r) \sim \mathcal{N}\left(\hat{\theta}_i(m, r), s_i\right)$. Using this approximation, we can obtain an estimate of the mean SampEn MSE for all $n$ signals by sampling from its respective distribution. Algorithm \ref{alg:gaussian-signal-signal-mse-approx} details the approach.

\begin{algorithm*}
\caption{Gaussian Approximation of Signal Set SampEn Estimate MSE}
\begin{algorithmic}[1]
\Require $\boldsymbol{S} = \left\{\boldsymbol{x}_1, \ldots, \boldsymbol{x}_n \right\}$, $m \in \mathbb{Z}_+$, $r \in \mathbb{R}_+$, $D \in \mathbb{Z}_+$, $\lambda \geq 0$
\For{$i = 1, \ldots, n$}
    \State Compute SampEn estimate of $\boldsymbol{x}_i$ given $(m, r)$: $\hat{\theta}_i(m, r)$.
    \State Calculate the SampEn estimate standard error, $s_i$ using Eq. (\ref{eq:sampen-se-approx}).
    \State $\left\{\widetilde{\theta}_{i, d}(m, r) \right\}_{d=1}^D \sim \mathcal{N}\left(\hat{\theta}_i(m, r), s_i \right)$.
    \State $\widehat{\epsilon}_i(m, r) = \frac{1}{D}\sum_{d=1}^D \left(\widetilde{\theta}_{i, d}(m, r) - \hat{\theta}_i(m, r) \right)^2$.
\EndFor
\Return $\frac{1}{n} \sum_{i=1}^n \widehat{\epsilon}_i(m, r) + \lambda\sqrt{r}$
\end{algorithmic}
\label{alg:gaussian-signal-signal-mse-approx}
\end{algorithm*}

Using this estimate, we can obtain an approximate comparison of the objective value realized for an $(m, r)$ combination selected by an alternate approach compared to our proposed BO algorithm.

\section{Constructing Weakly Stationary Signal Sets}
\label{appendix:weakly-stationary-signals}

Let $\boldsymbol{S}_j = \left\{\boldsymbol{x}_1, \ldots, \boldsymbol{x}_{n_j} \right\}$ where $\boldsymbol{x}_i \in \mathbb{R}^{N_j}$ correspond to the $j$-th signal set analyzed in Section \ref{subsection:real-experiments}. To ensure valid SampEn analysis \citep{chatain2019effects}, we require that all signals contained in the signal set are weakly stationary. We evaluate their statistical stationarity via the Augmented Dickey-Fuller (ADF) test \citep{dickey1979distribution} and corrected for multiple testing error at confidence level $\alpha = 0.05$ using the Holm-Sidak method \citep{holm1979simple, vsidak1967rectangular}. Algorithm \ref{alg:building-weakly-stationary-signal-sets} summarizes this process for a particular signal set.

\begin{algorithm*}[htbp]
\caption{Weakly Stationary Signal Set Construction}
\begin{algorithmic}[1]
\Require $\boldsymbol{S} = \left\{\boldsymbol{x}_1, \ldots, \boldsymbol{x}_n \right\}$, $\alpha \in (0, 1)$
\State $\widetilde{\boldsymbol{S}} \gets \emptyset$ \Comment{Placeholder for modified signal set}
\For{$i = 1, \ldots, n$}
    \State Calculate differenced signal: $\widetilde{\boldsymbol{x}}_i = \left(x_2 - x_1, x_3 - x_2, \ldots, x_N - x_{N-1} \right)$
    \State Normalize $\widetilde{\boldsymbol{x}}_i$ to have zero mean and unit variance
    \State Compute ADF test statistic p-value, $p_i$ \Comment{See \citep{dickey1979distribution}}
    \State $\widetilde{\boldsymbol{S}} \gets \widetilde{\boldsymbol{S}} \cup \left\{\widetilde{\boldsymbol{x}}_i \right\}$
\EndFor
\State Given the $n$ ADF p-values: $\left\{p_i \right\}_{i=1}^n$, compute the multiple testing error adjusted p-values, $\left\{\widetilde{p}_i \right\}_{i=1}^n$ using the Holm-Sidak method \Comment{See \citep{holm1979simple, vsidak1967rectangular}}
\For{$i = 1, \ldots, n$}
    \If{$\widetilde{p}_i > \alpha$}
        \State Remove $\widetilde{\boldsymbol{x}}_i$ from $\widetilde{\boldsymbol{S}}$
    \EndIf
\EndFor
\Return $\widetilde{\boldsymbol{S}}$
\end{algorithmic}
\label{alg:building-weakly-stationary-signal-sets}
\end{algorithm*}

\end{document}